%% file: main.tex
\documentclass[aps,reprint,amsmath,amssymb, aps,pre,floatfix,superscriptaddress]{revtex4-1}
\usepackage{latexsym,amssymb,amsfonts,mathrsfs,amsmath,bm}
\usepackage[linktocpage,colorlinks=true,linkcolor=blue,citecolor=blue,breaklinks=true,urlcolor=blue]{hyperref}
\usepackage{graphicx}
\usepackage{siunitx}
\usepackage{soul}
\usepackage{xcolor}
\usepackage[version=4]{mhchem}
\usepackage{tikz}
\usepackage{scalerel}
\usetikzlibrary{svg.path}

\renewcommand{\vec}{\bm}

\definecolor{orcidlogocol}{HTML}{A6CE39}
\tikzset{
  orcidlogo/.pic={
    \fill[orcidlogocol] svg{M256,128c0,70.7-57.3,128-128,128C57.3,256,0,198.7,0,128C0,57.3,57.3,0,128,0C198.7,0,256,57.3,256,128z};
    \fill[white] svg{M86.3,186.2H70.9V79.1h15.4v48.4V186.2z}
                 svg{M108.9,79.1h41.6c39.6,0,57,28.3,57,53.6c0,27.5-21.5,53.6-56.8,53.6h-41.8V79.1z M124.3,172.4h24.5c34.9,0,42.9-26.5,42.9-39.7c0-21.5-13.7-39.7-43.7-39.7h-23.7V172.4z}
                 svg{M88.7,56.8c0,5.5-4.5,10.1-10.1,10.1c-5.6,0-10.1-4.6-10.1-10.1c0-5.6,4.5-10.1,10.1-10.1C84.2,46.7,88.7,51.3,88.7,56.8z};
  }
}

\newcommand\orcid[1]{\href{https://orcid.org/#1}{\mbox{\scalerel*{
\begin{tikzpicture}[yscale=-1,transform shape]
\pic{orcidlogo};
\end{tikzpicture}
}{|}}}}

\usepackage[utf8]{inputenc}
\begin{document}
\title{Collective entrainment and confinement amplify transport by schooling micro-swimmers}

\author{Chenyu Jin\orcid{0000-0002-5552-0340}}
\affiliation{Experimentalphysik I, Universität Bayreuth, Bayreuth, Germany}
\affiliation{Max Planck Institute for Dynamics and Self-Organization and Institute for the Dynamics of Complex Systems, Georg August Universit\"at, 37077 G\"ottingen, Germany}
\author{Yibo Chen\orcid{0000-0001-6786-707X}} 
\affiliation{Physics of Fluids Group, Max Planck Center for Complex Fluid Dynamics, MESA+ Institute and J. M. Burgers Center for Fluid Dynamics, University of Twente, PO Box 217,7500 AE Enschede, The Netherlands}
\author{Corinna C. Maass\orcid{0000-0001-6287-4107}}
\email{c.c.maass@utwente.nl}
\affiliation{Max Planck Institute for Dynamics and Self-Organization and Institute for the Dynamics of Complex Systems, Georg August Universit\"at, 37077 G\"ottingen, Germany}
\affiliation{Physics of Fluids Group, Max Planck Center for Complex Fluid Dynamics, MESA+ Institute and J. M. Burgers Center for Fluid Dynamics, University of Twente, PO Box 217,7500 AE Enschede, The Netherlands}

\author{Arnold J. T. M. Mathijssen\orcid{0000-0002-9577-8928}}
\email{amaths@upenn.edu}
\affiliation{Department of Physics \& Astronomy, University of Pennsylvania, 209 South 33rd Street, Philadelphia, PA 19104, USA}

\date{\today}

\begin{abstract}
Micro-swimmers can serve as cargo carriers that move deep inside complex flow networks. 
When a school collectively entrains the surrounding fluid, their transport capacity can be enhanced. 
This effect is quantified with good agreement between experiments with self-propelled droplets and a confined Brinkman squirmer model. 
The volume of liquid entrained can be much larger than the droplet itself, amplifying the effective cargo capacity over an order of magnitude, even for dilute schools. 
Hence, biological and engineered swimmers can efficiently transport materials into confined environments.
\end{abstract}

\maketitle

\section*{Introduction}

``At low Reynolds number you can't shake off your environment. If you move, you take it along''  \cite{purcell1977life}. 
This principle, that a moving particle permanently displaces its surrounding fluid, was quantified in 1953 by Darwin \cite{darwin1953note, eames1994drift}. 
The amount of liquid entrained, the `Darwin drift volume', diverges for a colloid pulled along an infinite straight line through an unconfined Stokesian fluid \cite{eames2003fluid, shaik2020drag}, also at intermediate Reynolds numbers \cite{chisholm2017drift}. 
However, the drift volume is finite for micro-swimmers that do not exert a net force on the liquid \cite{pushkin2012fluid, chisholm2018partial}.
Hydrodynamic entrainment can be a curse and a blessing: It is important in a wide range of biological and ecological processes, including enhanced diffusion \cite{wu2000particle, kim2004enhanced, leptos2009_dynamics, thiffeault2010stirring, lin2011_stirring, pushkin2013fluid, morozov2014enhanced, peng2016diffusion, jeanneret2016_entrainment, mathijssen2021active}, biogenic mixing \cite{katija2009viscosity, subramanian2010viscosity, nawroth2014induced, houghton2018vertically, mathijssen2019collective, ortlieb2019statistics}, food uptake \cite{magar2003nutrient, short2006flows, michelin2011optimal, tam2011optimal, mathijssen2019nutrient}, particle transport \cite{papavassiliou2015many, shum2017entrainment, mueller2017transport, vaccari2018cargo, purushothaman2021hydrodynamic}, fungal spore dispersal \cite{ingham2011mutually}, oxygen redistribution \cite{tuval2005bacterial}, and microbial interaction probabilities \cite{mathijssen2018_universal}.

Here, we consider collective entrainment by a school (flock, or swarm) of micro-swimmers. 
Beyond biological systems this could equally be advantageous for micro-robotic material transport, especially in confined environments. 
A wide range of synthetic micro-swimmers has been developed in recent years, including active Janus colloids, magnetic swimmers, and bimetallic nanorods \cite{dreyfus2005microscopic, howse2007selfmotile, peng2014induced, nishiguchi2015mesoscopic, bechinger2016active, elgeti2015physics, zottl2016emergent, campbell2019_experimental}.
While these swimmers could serve as carriers that transport a payload deep into a network of micro-channels, their internal cargo space is inherently limited \cite{nelson2010microrobots, tiwari2012drug, wilhelm2016analysis, shen2017high, xu2019self, alapan2019microrobotics, erkoc2019mobile, ider2020tuning, yang2020motion}.
Therefore, instead of transporting cargo inside the micro-swimmer, we consider entraining the medium outside it, so that passive cargo vesicles or compounds dissolved in the fluid itself can be pushed forwards by groups of swimmers cooperatively. 
To investigate this quantitatively, we focus on self-propelled droplets \cite{thutupalli2011swarming, maass2016_swimming}, which feature a well-developed toolbox for tuning their motility, flow generation, collective dynamics, and solving microfluidic mazes \cite{thutupalli2013tuning, izri2014self, krueger2016dimensionality, krueger2017curling, jin2017chemotaxis, thutupalli2018flow, jin2018chemotactic, blois2019flow, stamatopoulos2020droplet, hokmabad2021_emergence, lippera2020collisions, hokmabad2020quantitative}.

\begin{figure}[b]
    \includegraphics[width=\linewidth]{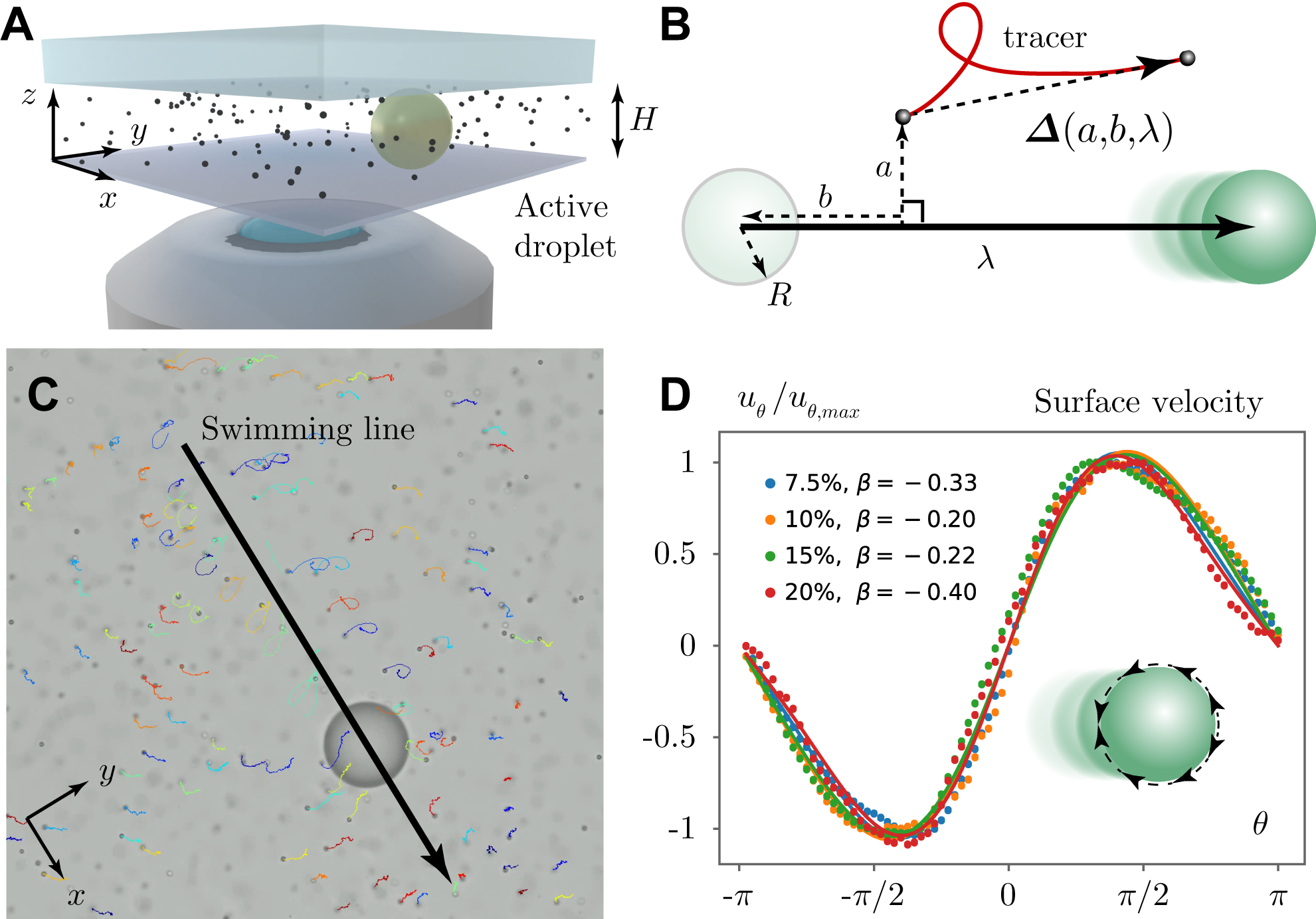}
    \caption{
    Fluid transport by a self-propelled droplet.
    \textbf{(A)} Schematic of experimental setup. Cell bottom glass, top PDMS, $H=\SI{52}{\um}$, droplet radius $R=\SI{24}{\um}$.
    \textbf{(B)} Diagram of an entrainment event. Due to the flow generated by the droplet, the tracer is displaced by $\vec{\Delta}(a,b,\lambda)$ as a function of the impact parameters $a$ and $b$, and the swimming path length $\lambda$.
    \textbf{(C)} Movie screenshot showing particle trajectories due to an active droplet swimming along a straight line [see Video 1].
    \textbf{(D)} Tangential flow velocity at the surface of the droplet, in the co-moving frame of reference. Data points are PIV measurements close to the interface, for 4 different surfactant concentrations. Lines are fits to the first two modes of the squirmer model [Eq.~\ref{eq:SquirmerSurfaceVelocity}]. Legend: Resulting dipole coefficient $\beta=B_2/B_1$ for each TTAB concentration.
    }
    \label{fig1}
\end{figure}

\subsection*{Experimental setup}
%

\begin{figure}[t!]
    \includegraphics[width=\linewidth]{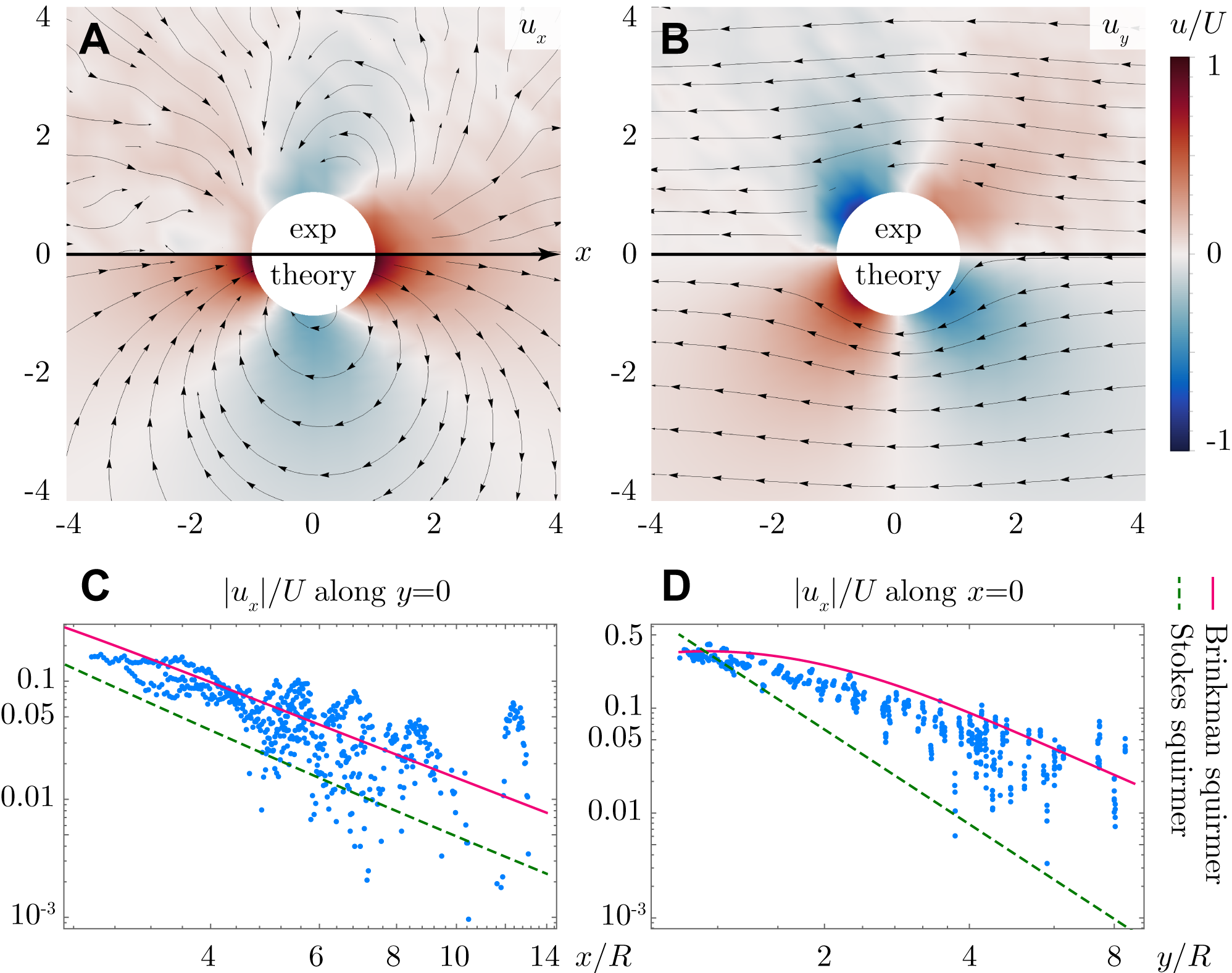}
    \caption{
    Flows $\vec{u}(\vec{r})$ generated by an active droplet.
    \textbf{(A,B)} Comparison of the flow fields between (top half) experiments and (bottom half) the squirmer model in a Brinkman fluid [Eq.~S13]. 
    Flows are shown at the mid plane of the microfluidic chamber.
    Streamlines are shown (A) in the laboratory frame, and (B) in the frame co-moving with the droplet.
    Background colours denote (A) the longitudinal flow $u_x$ and (B) the transverse flow $u_y$, scaled with the swimming speed. Note that the transverse flow $u_y$ changes sign across the $x$ axis.
    \textbf{(C,D)} Decay of the longitudinal flow strength in the laboratory frame along the (C) $x$-axis and (D) $y$-axis. 
    Points show experiments and the solid line is the corresponding squirmer model in a Brinkman fluid. The dashed line shows the squirmer model in an unconfined Stokesian fluid, which underestimates the flow strengths.
    }
    \label{fig2}
\end{figure}

We begin by collecting data on the dynamics of tracer particles displaced by self-propelling droplets.
Our experimental methods are described in \S\,I of the Supplemental Material (SM) 
\footnote{See Supplemental Material containing experimental methods, theoretical derivations, a supplementary figure, 3 videos, and additional references \cite{qin2010_soft, thorsen2001_dynamic, crocker1996_methods, Jeanneret2019, tsay1991viscous, pepper2010_nearby, Pushkin2016, nganguia2018squirming, nganguia2020squirming, whitaker1986flow, gilpin2017vortex, liu2016bimetallic, weeks1971role, eames1999connection, dabiri2005estimation}.}.

Briefly, in our set-up [Fig.~\ref{fig1}A] we place a CB15 oil droplet of radius $R=\SI{24}{\micro\metre}$ in quasi-2D confinement of a polydimethylsiloxane (PDMS)-glass reservoir with $H=\SI{52}{\micro\metre}$, where it self-propels with a speed of $U = \SI[separate-uncertainty=true]{21.8\pm0.5}{\micro\metre \per\second}$.
The swimming medium is a micellar aqueous solution of TTAB surfactant at 5\% mass fraction, containing a low density ($\approx 10^{5}/\mu$l) of polystyrene tracer colloids with diameter $d_C=\SI{2}{\micro\metre}$. 
Heavy water is added to the swimming medium in order to match its density with the tracer colloids.
The droplet propulsion is powered by a self-supporting interfacial surface tension gradient~\cite{herminghaus2014interfacial,maass2016_swimming}.
Using video microscopy, we extract trajectories for approximately $300$--$400$ tracer colloids [Fig.~\ref{fig1}B,C \& Video 1] whose $z$ position did not deviate too far from the chamber's mid plane during the transit of the active droplet.
After correcting a slight background drift, all coordinates are translated and rotated such that the swimmer moves along the $x$ axis.
Hence, we determine the displacement of a tracer particle, $\vec{\Delta}(a,b,\lambda)$, following the notation by Lin \textit{et al.} \cite{lin2011_stirring}, where the impact parameters $a, b$ specify the initial perpendicular and parallel distance of the tracer to the start of the swimming path, and $\lambda$ is the swimming path length [Fig.~\ref{fig1}B].
The droplets move persistently, with $\lambda\approx 100R$ before they reorient due to rotational fluctuations, likely from inhomogeneities in the surfactant coverage.

\subsection*{Flow generation}
%
We then seek to describe the flow $\vec{u}(\vec{r})$ produced by the droplet due to its self-propulsion.
Using particle image velocimetry (PIV), we first measure the tangential flow velocity $u_\theta$ near the droplet interface [Fig.~\ref{fig1}D; points]. 
These experiments are compared with the squirmer model \cite{lighthill1952squirming, blake1971spherical}, where the surface velocity is given by the mode decomposition,

    \begin{align}
    \label{eq:SquirmerSurfaceVelocity}
    \left. u_\theta \right|_{r=R} = B_1\sin(\theta)+ \frac{1}{2} B_2 \sin(2\theta) + \dots
    \end{align}
By fitting this model to our data, we find the mode strengths $B_1 = \SI[separate-uncertainty=true]{29.8 \pm 2}{\micro\metre \per\second}$ and $B_2 = \SI[separate-uncertainty=true]{-7.8 \pm 2}{\micro\metre \per\second}$
Using only these first two modes already offers a good agreement with the measured flow [Fig.~\ref{fig1}D; lines].
The dipole coefficient is $\beta = B_2/B_1 \approx -0.3$, so the droplets are weak pushers.
These experiments are repeated for 4 different TTAB concentrations [Fig.~\ref{fig1}D; legend].

The measured flows generated by an active droplet are shown in the top halves of Fig.~\ref{fig2}A,B.
These experiments agree well with the results by De Blois \textit{et al.} \cite{blois2019flow}, showing strong regions of forward flow [red colours in panel A].
This fluid motion leads to entrainment, pushing particles in front of the droplet or dragging them along behind.
Quantitatively, however, the measured flow [Fig.~\ref{fig2}C,D; blue points] is about 5 times stronger than the prediction from the conventional squirmer model in a 3D unconfined Stokesian fluid [green dashed lines]. 

To provide a more accurate theoretical description of the flows created by an active droplet [SM \S\,II], a squirmer model was developed that accounts for the quasi-2D confinement using the Brinkman equations \cite{brinkman1947_calculation}.
The quintessential difference between the conventional Stokes model and the Brinkman model is the confinement, which is described by a permeability parameter in terms of the microfluidic chamber height, $k = 12/H^2$, analogous to Darcy's law.
The Brinkman squirmer model is then derived by imposing the same tangential velocity at the droplet interface [Eq.~\ref{eq:SquirmerSurfaceVelocity}] as boundary condition.
The resulting expression [Eq.~S13] gives an accurate description of the measured flows in all directions [Fig.~\ref{fig2}A,B; bottom halves].
Potential sources of experimental error are tracers having a small velocity component out of the focal plane, the difficulty of sampling near the droplet interface, and the droplet diameter being a bit smaller than the channel height. 
Therefore, the mean experimental flows are slightly weaker than modelled. 
Still, compared to the squirmer model in an unconfined Stokesian fluid, the Brinkman model offers a significantly improved agreement [Fig.~\ref{fig2}C,D; pink lines].

\subsection*{Particle entrainment}
%
Having quantified the flows made by active droplets, we consider how these currents displace tracer particles (or equivalently, the fluid itself) along the swimming direction.
The tracers' equation of motion is

    \begin{align}
    \label{eq:ParticleEOM}
    \frac{d \vec{r}_\text{T}}{dt} &= \vec{u}(\vec{r}_\text{T} - \vec{r}_\text{S}) + \sqrt{2D}\vec{\xi}(t),
    \end{align}
where $\vec{r}_\text{S}(t)$ is the moving droplet position, the flow $\vec{u}$ is given by Eq.~S13, $\vec{\xi}$ is standard white noise, and the particle diffusivity, $D=\SI{0.22}{\micro\metre^2\per\second}$, is determined experimentally by analysing their Brownian motion in the absence of droplets.
The particle displacement is then defined as $\vec{\Delta}=\vec{r}_\text{T}(t_\text{f}) - \vec{r}_\text{T}(t_\text{i})$, where $t_\text{i}=0$ and $t_\text{f}=\lambda/U$. 

\begin{figure}[t!]
    \includegraphics[width=\linewidth]{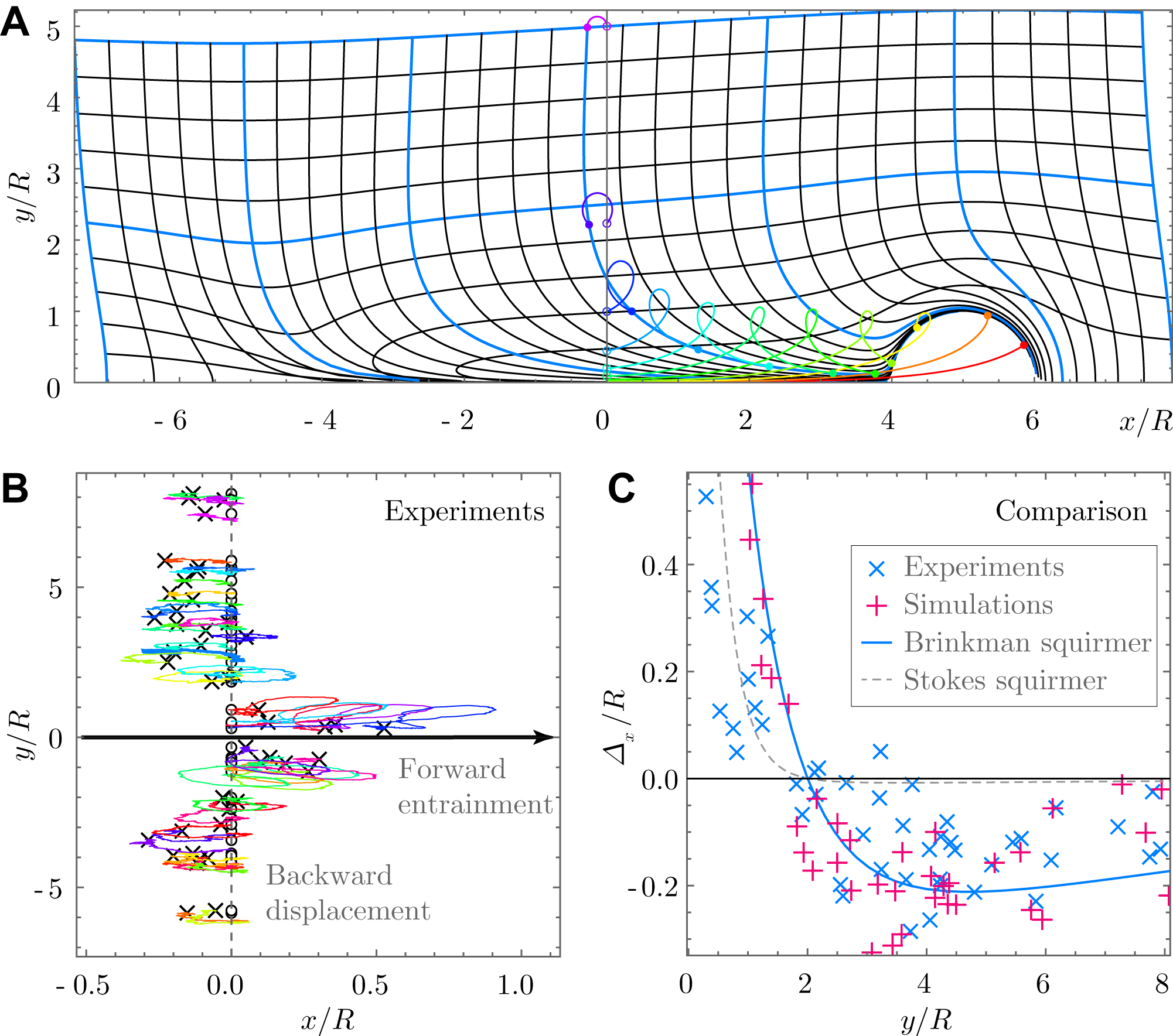}
    \caption{
    Entrainment of particles by a confined self-propelled droplet.
    \textbf{(A)} Fluid transport by a squirmer in a Brinkman fluid. Shown is the deformation of an initially uniform rectangular grid of tracer particles, after a droplet swims from $x/R=-5$ to $5$. Coloured trajectories show the motion of particles starting at different positions along the $y$ axis.
    \textbf{(B)} Experimental trajectories, with their initial positions aligned along the $y$ axis. The initial positions are marked with a circle ($\circ$) and the final positions are marked with a cross ($\times$). 
    \textbf{(C)} Comparison. Blue crosses ($\times$) show the experimental entrainment along $x$ as a function of lateral distance $y$, pink plus signs ($+$) show simulated trajectories using the same initial positions using the Brinkman model with Brownian motion, the solid blue line shows the Brinkman theory without noise, and the dashed grey line shows the squirmer model in an unconfined Stokesian fluid. Far away the particles are displaced backwards, but nearby they are entrained forwards, along the swimming direction.
    }
    \label{fig3}
\end{figure}

To illustrate the impact of fluid deformation by a self-propelled droplet, following Pushkin \textit{et al.} \cite{pushkin2012fluid}, we first simulate the Brinkman model in the absence of noise.
Initially, the fluid parcels are arranged along a square grid, after which the droplet swims through [Fig.~\ref{fig3}A \& Video 2].
The horizontal curtains (initially same $y$ values) are pushed outwards in front of the drop and pulled inwards behind it.
The vertical curtains are folded around the droplet, compressed in front and stretched out behind the drop. 
In the middle of the swimming path, $x=0$, particles are displaced backwards for large $y$ values but nearby the particles are strongly entrained forwards [Trajectories; from violet to red].

The same is observed experimentally [Fig.~\ref{fig3}B]. 
To quantify the entrainment, we select particles that are initially located close to the middle of the swimming path, with $\lambda = 15R$ and impact parameter $5<b/R<10$.
Far from the droplet, for impact parameters $a \gtrsim 2R$, the tracers have a final displacement that is backwards compared to the swimming direction.
Note that it is not always possible to track particles close to the drop because they can move out of the focal plane.
Nevertheless, for $a \lesssim 2R$, we still see a strong forward entrainment.

These measurements are in good agreement with the Brinkman squirmer model [Fig.~\ref{fig3}C].
Here we simulate particles with Brownian motion [Eq.~\ref{eq:ParticleEOM}], where the computed trajectories start from the same initial positions as the experimental particles.
Interestingly, these results do not deviate significantly from the deterministic Brinkman model (solid line), obtained from numerically integrating Eq.~\ref{eq:ParticleEOM} without Brownian motion and using the same droplet dynamics as in the experiment, which may also be approximated theoretically using asymptotic analysis \cite{mathijssen2015_tracer}.
However, the transport of particles is substantially different in bulk liquids (dashed line) compared to confined spaces (solid line).

\subsection*{Collective transport}
%

\begin{figure}[t!]
    \includegraphics[width=\linewidth]{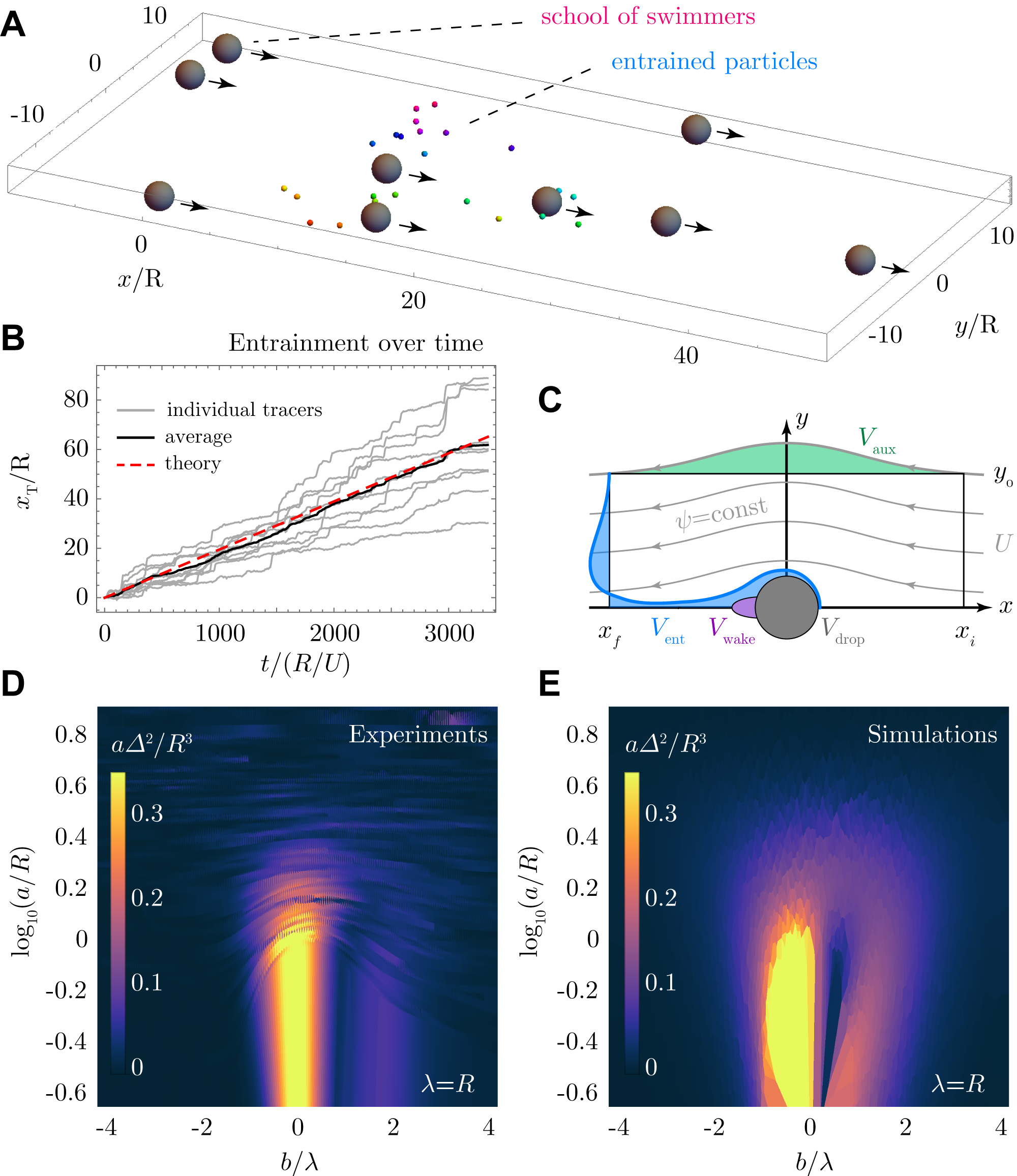}
    \caption{
    Entrainment by a school of micro-swimmers.
    \textbf{(A)} 
    Simulation snapshot; see Video 3. Self-propelled droplets (black) move collectively along the positive $x$ direction, transporting particles (colours) that initially started at $x=0$.
    \textbf{(B)} 
    Displacement of the particles along the swimming direction as a function of time, $x_\text{T}(t)$. Individual trajectories (grey lines) show jumps and drift events, which on average (black line) lead to the entrainment velocity $U_\text{ent}$ that is predicted analytically (red dashed line) by Eq.~\ref{eq:TransportVelocity}.
    \textbf{(C)} 
    Diagram of the entrainment volume (blue shaded area) that is dragged behind an active droplet, in the co-moving reference frame. The grey lines depict stream lines, the purple area is the swimmer wake, and the green area is the auxiliary volume.
    \textbf{(D,E)} 
    Stirring by active droplets in a Brinkman fluid.
    Shown is the integrand $aR^{-3}\Delta^2(a,b,\lambda)$ from Eq.~\ref{eq:EffectiveDiffusivity1}, measured experimentally and computed numerically.
    }
    \label{fig4}
\end{figure}

Next, we consider collective transport by a school of micro-swimmers, where particles or fluid parcels are pushed forwards by multiple subsequent entrainment events [Fig.~\ref{fig4}A \& Video 3].
This principle could apply to a broad range of biological and synthetic micro-swimmers.
Here, we focus on drops that are distributed uniformly in the $xy$ plane with area fraction $\phi$, so the number density is $n=\phi/\pi R^2$, and all droplets swim in the $x$ direction with speed $U$.
Depending on the kind of micro-swimmer, this orientation alignment could stem from collective interactions \cite{krueger2016dimensionality, thutupalli2018flow}, internal mechanisms like chemotaxis \cite{jin2017chemotaxis}, or guidance by tracks in the microfluidic channel \cite{simmchen2016topographical, stamatopoulos2020droplet}. 
Thus, having characterised the entrainment due to a single droplet in the last section, we now simulate tracers that are entrained by a school [see SM \S\,III]. 

As shown in Fig.~\ref{fig4}B, the individual tracers [grey trajectories] experience large sudden displacement events when a droplet transits nearby (jumps), which are separated by long periods of comparatively weak Brownian motion and long-ranged flows (drifts).
Averaged over time, or equivalently over a statistical ensemble, these jumps and drifts give rise to a mean entrainment velocity, 

    \begin{align}
    \label{eq:EntrainmentVelocity}
    \vec{U}_\text{ent} = \left \langle  \frac{d \vec{r}_\text{T}  }{dt} \right \rangle,
    \end{align}
which quantifies how fast the fluid (and all the particles it contains) is transported along the swimming direction [black solid line].

This mean entrainment velocity can be approximated analytically.
For a dilute school of swimmers, $\phi \ll 1$, we can write $U_\text{ent} = n U A_\text{ent}$, where the first two factors encode the encounter rate, and $V_\text{ent} = H A_\text{ent}$ is the volume of liquid entrained by a single droplet.
This is also called the ``Darwin drift'' volume \cite{darwin1953note, pushkin2012fluid}, the region swept out by a tracer curtain [Fig.~\ref{fig4}C; blue region].
Specifically, in our quasi-2D geometry and in the limit of $\lambda \to \infty$ and $b=\lambda/2$, the entrainment volume is defined as 

    \begin{align}
    \label{eq:EntrainmentVolume2D}
    V_\text{ent} &= 2H\int_0^\infty  \Delta_x da.
    \end{align}
To compute this quantity analytically [SM \S\,III B], we use a streamfunction formulation to determine the Darwin drift volume in the confined Brinkman medium.
Hence, we find that a school of squirmers can transport their surrounding fluid with velocity

    \begin{align}
    \label{eq:TransportVelocity}
    \frac{ U_\text{ent}}{U}
    &=  \frac{\phi V_\text{ent}}{\pi R^2 H}
    = \phi  \left(1 + \frac{2 \left(2 U-B_1\right) h K_1\left(\frac{R}{h }\right)}{U R K_0\left(\frac{R}{h }\right)}\right),
     \end{align}
which applies to both pushers and pullers with dipole coefficient $|\beta|<1$. 
Here $h=\frac{H}{\sqrt{12}}$, and $K_n(x)$ are modified Bessel functions of the second kind.
Inserting $\phi=0.01$ and our experimental values into Eq.~\ref{eq:TransportVelocity}, 
we obtain the red dashed line in Fig.~\ref{fig4}B, which offers a good agreement with the simulation.
The school can generate a significant fluid transport, because all the liquid between the swimmers also moves, with speed $U_\text{ent}$ on average.
Therefore, a substantial amount of cargo can be moved within the medium, even if $U_\text{ent}$ is small compared to $U$.
To quantify this, we write the internal transport as $\Theta_0 = \phi U \eta$, where $\eta \in [0,1]$ denotes the internal storage efficiency, defined as the volume of the cargo held compared to the total volume of the swimmer.
In drug delivery this quantity is also called the loading capacity, which is typically $\eta \sim 0.1$ or less \cite{nelson2010microrobots, tiwari2012drug, wilhelm2016analysis, shen2017high, xu2019self, alapan2019microrobotics, erkoc2019mobile, ider2020tuning, yang2020motion}.
Then, the total transport is

    \begin{align}
    \label{eq:CargoCapacity}
    \Theta = \phi U \eta + (1-\phi)U_\text{ent}.
    \end{align}
For an ideal carrier, $\eta=1$, using $\phi=0.01$ and $U_\text{ent}\approx0.02U$ from Eq.~\ref{eq:TransportVelocity}, the total transport is already tripled by entrainment, $\Theta/\Theta_0 \approx 3$.
Moreover, if the space for cargo inside the swimmer is poor, $\eta=0.1$, the total transport is over twenty times enhanced, $\Theta/\Theta_0 \approx 20$.

Interestingly, the confinement can help with the particle transport.
For a squirmer in an unconfined Stokesian fluid, Pushkin \textit{et al.}~\cite{pushkin2012fluid} found that entrainment volume is equal to half the droplet volume, $V_\text{ent}/V_\text{drop} = 0.5$, if $|\beta|<1$.
In the confined Brinkman medium, we find $V_\text{ent}/V_\text{drop} \approx 2.93$ [Eq.~S24], so about six times larger.
This result is confirmed by integrating the curtain displacement [Eq.~\ref{eq:EntrainmentVolume2D}] numerically [SM \S\,III C].

\subsection*{Enhanced mixing}
%
Before closing this article, we also discuss how swimming droplets can enhance particle diffusion \cite{wu2000particle, kim2004enhanced, leptos2009_dynamics, thiffeault2010stirring, lin2011_stirring, pushkin2013fluid, morozov2014enhanced, peng2016diffusion, jeanneret2016_entrainment}.
Building on the same framework, we now consider droplets without alignment that perform uncorrelated, random reorientations in the $xy$ plane, with swimming path length $\lambda$.
Then, the total displacement $\Delta(a, b, \lambda)$ can be used to estimate the stirring efficiency. 
Combining equations 2.5 and 3.2 by Lin \textit{et al.}~\cite{lin2011_stirring}, for a dilute swimmer suspension, the enhanced diffusivity is given by
    \begin{equation}
    \label{eq:EffectiveDiffusivity1}
    D_\text{enh}
    =
    \frac{\phi R U}{2\pi} \int_{\mathbb{R}^2} \frac{a \Delta^{2}(a, b, \lambda)}{R^3} ~\mathrm{d} \left ( \frac{b}{\lambda} \right) ~\mathrm{d} \log \left( \frac{a}{R} \right).
    \end{equation}
Using our particle tracking experiments, we measure the integrand of this expression as a function of the impact parameters $a$ and $b$ [Fig.~\ref{fig4}D].
These findings compare favourably with the simulated values from the Brinkman squirmer model [Fig.~\ref{fig4}E].
Hence, by computing the integrand for $a/R\in [10^{-2}, 10^3]$ and $b/\lambda\in [-10, 10]$, using $\lambda=100R$, and integrating these results numerically, we find that the enhanced diffusion is $D_\text{enh} / \phi \approx \SI{638}{\micro\metre^2\per\second}$.
Hence, the mixing efficiency can be significantly larger compared to thermal diffusion of micron-sized particles ($D_\text{th} \approx \SI{0.2}{\micro\metre^2\per\second}$), even at low droplet densities.
We can also compare between active droplets in a quasi-2D fluid and in an unconfined 3D fluid, using the results by \citet{lin2011_stirring}. Inserting the parameters of our droplets into their figure 6a expression, one finds $D_\text{enh}/\phi \sim 0.5 nUR^4/\phi \sim \SI{62}{\micro\metre^2 \per \second}$ in an unconfined fluid. That is about an order of magnitude lower than the quasi-2D result, so the confinement can also amplify particle mixing.

\subsection*{Discussion}
%
To conclude, we investigated the transport of fluid (or particles embedded therein) by self-propelled droplets.
First, by measuring the flows they generate, a Brinkman squirmer model was developed that accounts for the quasi-2D confined geometry in a tractable manner.
We then quantified the amount of liquid entrained by a single droplet, and we derived the corresponding Darwin drift volume analytically.
To reveal how fast a school of droplets can push cargo particles through a microfluidic channel, the transport velocity was computed by integrating over successive entrainment events.
Hence, we found that the total cargo capacity can be enhanced significantly compared to transport inside the micro-carriers alone.
In that sense, we call this phenomenon `collective entrainment', because it relies on the cooperation of a large collection of swimmers, together pushing the cargo forward. The single-swimmer Darwin drift volume already captures the main physics for the average transport velocity in the dilute limit, but the collective transport only arises when the swimmers align with one another in a school.

While this paper focuses on self-propelled droplets to compare our experiments with the theory, we envisage that these results are widely applicable to other types of micro-swimmers, particularly active Janus particles \cite{dreyfus2005microscopic, howse2007selfmotile, peng2014induced, nishiguchi2015mesoscopic, bechinger2016active, elgeti2015physics, zottl2016emergent, campbell2019_experimental}. The dipole moment could be large for such swimmers, which changes the entrainment volume. Therefore, we extend our discussion of Eq.~\ref{eq:TransportVelocity} for the case of $|\beta|>1$ in SM~\S\,III E.
Interestingly, the entrainment volume is the same for pushers and pullers [Fig.~S1].
More generally, using the universality of this model, one could account for almost any hydrodynamic signature with higher-order moments.
Thus, together with recent developments in micro-robotics \cite{xu2019self, yang2020motion}, an efficient cargo transport could be established in microfluidic networks or drug delivery applications \cite{nelson2010microrobots, tiwari2012drug, wilhelm2016analysis, shen2017high, xu2019self, alapan2019microrobotics, erkoc2019mobile, ider2020tuning, yang2020motion}.

Besides synthetic active particles, these results could be equally significant for material transport by biological micro-swimmers \cite{jeanneret2016_entrainment, ortlieb2019statistics}.
By directed collective motion through confined spaces \cite{zottl2014hydrodynamics, wioland2016directed}, they could entrain nutrients or signalling molecules deep into complex networks, including biofilms architectures \cite{vidakovic2018dynamic}, porous media \cite{bhattacharjee2019bacterial} or foams \cite{roveillo2020trapping}, much faster than transport by thermal diffusion.
Ultimately, self-propelled particles rely on the replenishment of fuel, particularly into confined spaces, to sustain their activity.

\subsection*{Acknowledgements}

AM acknowledges funding from the Human Frontier Science Program (Fellowship LT001670/2017) and the United States Department of Agriculture (USDA-NIFA AFRI grants 2020-67017-30776 and 2020-67015-32330). CCM and CJ acknowledge funding from the BMBF/MPG MaxSynBio initiative and the DFG SPP 1726 ``Microswimmers''.

\bibliography{entrainment.bib}
\cleardoublepage
\input{SIcontent.tex}

\end{document}

%% file: SIcontent.tex
\onecolumngrid
\section*{Supplemental Material: Collective entrainment and confinement amplify transport by schooling micro-swimmers}

 \renewcommand\thefigure{S\arabic{figure}} 
 \setcounter{figure}{0}
 \renewcommand\theequation{S\arabic{equation}} 
 \setcounter{equation}{0}

\section{Experimental methods}
\label{sec:Methods}

\subsection{Microfluidic chip fabrication}

The microfluidic devices are fabricated with standard soft lithography techniques \cite{qin2010_soft}. 
We design photomasks in a 2D AutoCad application and have them printed as a high-resolution emulsion film by an external company (128,000 dots per inch; JD Photo-Tools). 
From the printed photomask, we synthesize a SU-8 (Micro Resist Technology) photoresist structure on a Si wafer (Wafer World Inc.) in a clean room environment using UV lithography.
The mold is then used to generate PDMS (polydimethylsiloxane; Sylgard 184; Dow Corning) imprints. After punching in fluid inlets and outlets, the imprints are bonded to glass slides.
Covalent bonding between PDMS and glass is achieved by pretreating all surfaces in an air plasma (Pico P100-8; Diener Electronic GmbH + Co. KG) for $\SI{30}{\second}$.

\subsection{Droplet generation}

We produce monodisperse droplets of (S)-4-Cyano-4’-(2-615methylbutyl)biphenyl (CB15, Synthon Chemicals) in an aqueous solution of the cationic surfactant tetradecyltrimethylammonium bromide (TTAB, Sigma Aldrich, critical micellar concentration  $c_\mathrm{cmc}=0.13\,\text{~wt.\%}$) in flow-focusing microfluidic devices~\cite{thorsen2001_dynamic} at high production rates (\SIrange{10}{30}{\per\second}). Droplet radii $R$ can be tuned by flow rates and channel geometries, we aimed for $R=\SI{25}{\micro\metre}$.
The syringes (Braun) are mounted on a precision microfluidic pump (NEM-B101-02B; Cetoni GmbH) and connected to the inlets and outlets with Teflon tubing (39241; Novodirect GmbH). 

To generate oil-in-water emulsions, the primarily hydrophobic PDMS surfaces have to be hydrophilized by drawing a sequence of liquids through the outermost channel via a vacuum pump induced pressure gradient. 
The sequence constitutes: a 1:1 volumetric ratio of hydrochloric acid
(HCl at $37 \mathrm{wt.}\%$, Sigma-Aldrich) and hydrogen peroxide (\ce{H2O2} at $30 \mathrm{wt.}\%$, Sigma-Aldrich) for 2 minutes; Milli-Q water for 30 seconds; $5 \mathrm{wt.}\%$ aqueous solution of poly(diallyldimethylammonium chloride)
(PDADMAC, average molecular weight Mw $\approx 10^{5} \sim 2 \times 10^{5}$ g/mol , Sigma-Aldrich), for 2 minutes; $2 \mathrm{wt.}\%$ aqueous solution of poly(sodium 4-styrenesulphonate) (PSS, Mw $\approx 10^6$ g/mol , Sigma-Aldrich) for 2 minutes.

For droplet production and storage we use a submicellar $0.1\,\text{~wt.\%}$ TTAB solution. The concentration is sufficient to inhibit coalescence, but not high enough for oil solubilization or droplet swimming.

\subsection{Experimental protocol}
Experiments are carried out in quasi-2D PDMS-glass reservoirs fabricated using standard soft lithography protocols. The reservoirs are approximately $6$ mm $\times\, 10$ mm $\times\,  \SI{52}{\micro\metre}$ in size and contain arrays of 32 micropillars with radii $\SI{50}{\micro\metre}$ to support the ceiling.

At the beginning of each experiment,  we add $\approx{\SI{0.3}{\ul}}$ of stock oil emulsion to a \SI{10}{\ul}  $5\,\text{wt} \%$ TTAB solution (approximately $40\cdot c_\mathrm{cmc}$) containing a small amount of polystyrene tracer colloids ($d_\mathrm{C}=\SI{2}{\um}$) and pipette the mixture into the experimental container.

We observe the droplets on an Olympus IX-81 bright field microscope under 40x magnification. Images and movies are recorded by a commercial digital camera (Canon EOS 600D) at 25 frames per second.

\subsection{Data processing}

We process video microscopy data, tracking the swimmers / colloids and extracting trajectories, using software written in-house in Python/openCV, based on a Crocker-Grier type algorithm \cite{crocker1996_methods}. 
Owing to the significant size difference between droplet and the colloids, the video has to be processed twice for particle detection and tracking: First the droplet is tracked to generate a circular mask for each frame, which is then applied to block the droplet interior. The frames with the swimmer masked are then processed for colloid tracking.

There is often a slight background drift in the system, possibly due to residual flow after filling the reservoir. This background drift velocity is calculated by averaging the Brownian motion of the colloids outside of the entrainment volume. Before further analysis, this background drift is corrected for all the trajectories.

\section{Squirmer in a Brinkman medium}
\label{sec:BrinkmanSquirmer}

\subsection{Brinkman equations}

We consider a viscous fluid at low Reynolds number that follows the Stokes equations,
    \begin{equation}
    \label{eq:StokesEqn}
        \vec{\nabla} p = \mu \nabla^2 \vec{v}, \quad \vec{\nabla} \cdot \vec{v} = 0.
    \end{equation}
The fluid is confined in a channel of height $H$, with no-slip boundary conditions at the top and bottom surfaces. 
This confinement inherently changes the flow fields generated by micro-swimmers (see e.g. \cite{Jeanneret2019} and references therein).
To describe the confinement for active droplets, we use the thin film approximation by Brinkman and others \cite{brinkman1947_calculation, tsay1991viscous, pepper2010_nearby, Pushkin2016, nganguia2018squirming, nganguia2020squirming}.
Here we assume that the pressure is constant along the vertical $z$ direction, $p(x,y,z)=p(x,y)$, in Cartesian coordinates, and the flow velocity follows a Poiseuille profile, 
    \begin{equation}
        \vec{v}(x,y,z) = \vec{u}(x,y)\frac{6 z(H-z)}{H^2},
    \end{equation}
where the mean 2D flow is $\vec{u} = \int_0^H \vec{v} dz/H$.
Hence, we have $\frac{\partial^2}{\partial z^2} \vec{v} = - \frac{12}{H^2} \vec{u}$, so the Stokes equations (\ref{eq:StokesEqn}) can be approximated as
    \begin{equation}
    \label{eq:BrinkmanEqn}
        \vec{\nabla} p = \mu (\nabla^2 - k) \vec{u}, \quad \vec{\nabla} \cdot \vec{u} = 0,
    \end{equation}
which are called the Brinkman equations. 
Here the permeability $k = 12/H^2 = 1/h^2$, in direct analogy with Darcy's law \cite{whitaker1986flow}.

To solve the flows around an active droplet in this confined geometry, 
we use a polar coordinate system with the droplet centred at the origin.
The Brinkman equations can then be written as

    \begin{align}
    \label{eq:BrinkmanEqnPolar}
        \frac{1}{\mu} \frac{\partial p}{\partial r} 
        &=
        \frac{\partial}{\partial r}\left(\frac{1}{r} \frac{\partial}{\partial r}\left(r u_{r}\right)\right)+\frac{1}{r^{2}} \frac{\partial^{2} u_{r}}{\partial \theta^{2}}-\frac{2}{r^{2}} \frac{\partial u_{\theta}}{\partial \theta}- \frac{u_{r}}{h^2},
        \nonumber \\
        \frac{1}{\mu} \frac{1}{r} \frac{\partial p}{\partial \theta}
        &=
        \frac{\partial}{\partial r}\left(\frac{1}{r} \frac{\partial}{\partial r}\left(r u_{\theta}\right)\right)+\frac{1}{r^{2}} \frac{\partial^{2} u_{\theta}}{\partial \theta^{2}}+\frac{2}{r^{2}} \frac{\partial u_{r}}{\partial \theta}- \frac{u_{\theta}}{h^2},
        \nonumber \\
        0 & = \frac{1}{r} \frac{\partial}{\partial r}\left(r u_{r}\right)+\frac{1}{r} \frac{\partial u_{\theta}}{\partial \theta}. 
    \end{align}

\subsection{Surface actuation solution}
    
We first consider the surface actuation problem, also referred to as the pumping problem, where the droplet is stationary and drives tangential flows along its interface. 
As boundary conditions we then require that the flow vanishes as $r \to \infty$, that the radial velocity vanishes at the drop interface, $\left. u_r \right|_{r=R} = 0$, and the tangential velocity is decomposed into a series of modes:

    \begin{align}
    \label{eq:SquirmerModes1}
        \left. u_\theta \right|_{r=R} = \sum_{n=1}^\infty b_n \sin(n \theta).
    \end{align}
Therefore, we seek solutions of the form

    \begin{subequations}
    \begin{align}
    \label{eq:BrinkmanAnsatz}
        u_r(r, \theta)      &= U_r(r) \cos(n\theta),\\
        u_\theta(r, \theta) &= U_\theta(r) \sin(n\theta),\\
        p(r, \theta)        &= P(r) \cos(n\theta).
    \end{align}
    \end{subequations}
Inserting this ansatz into the Brinkman equations, we find the solution in terms of a streamfunction 
    \begin{equation}
    \label{eq:PumpingStreamfunction}
        \psi = \sum_{n=1}^\infty c_n R \left[ \left( \frac{R}{r} \right)^{n} -  \frac{K_n(r/h)}{K_n(R/h)} \right] \sin(n\theta),
    \end{equation}
where $K_n(x)$ is the modified Bessel function of the second kind of order $n$, and $K'_n(x)$ is its first derivative.
The connection between the prefactors $c_n$ and the mode strength $b_n$ of the tangential velocity at the surface is
    \begin{equation}
    \label{eq:ModeStrengths}
        c_n = - b_n \frac{h}{R}\frac{K_n(R /h)}{K_{n-1}(R /h)}.
    \end{equation}
Using the relations 
    \begin{equation}
    \label{eq:streamfunctionDerivatives}
        u_r = \frac{1}{r} \frac{\partial \psi}{\partial \theta}, \quad \quad
        u_\theta = -\frac{\partial \psi}{\partial r},
    \end{equation}
we find the velocity and pressure fields,

    \begin{subequations}
    \label{eq:PumpingSolutionBrinkmanSquirmer}
    \begin{align}
        u_r &= \sum_{n=1}^\infty c_n \left[ n \left( \frac{R}{r} \right)^{n+1} - n \frac{R}{r} \frac{K_n(r/h)}{K_n(R/h)} \right] \cos(n\theta),
        \\
        u_\theta &= \sum_{n=1}^\infty c_n \left[ n \left( \frac{R}{r} \right)^{n+1} + \frac{R}{h} \frac{K'_n(r/h)}{K_n(R/h)} \right] \sin(n\theta),
        \\
        p &= \sum_{n=1}^\infty c_n \left[ \frac{\mu R }{h^2} \left( \frac{R}{r} \right)^{n} \right] \cos(n\theta).
    \end{align}
    \end{subequations}
Note that in the squirmer literature, the surface velocity is sometimes also defined in terms of Legendre polynomials, 

    \begin{subequations}
    \begin{align}
    \label{eq:SquirmerModes2}
        \left. u_\theta \right|_{r=R} 
        &= \sum_{n=1}^\infty - B_n \frac{2}{n(n+1)} P^1_n(\cos \theta) 
        \\
        &= B_1 \sin(\theta) + \frac{1}{2} B_2\sin(2\theta)  + \dots  
    \end{align}
    \end{subequations}
This definition is related to Eq.~\ref{eq:ModeStrengths} via $B_1 = b_1$ and $B_2 = 2b_2$ for the first two modes. The dipole coefficient is then defined as $\beta = B_2/B_1 = 2b_2/b_1$.
    
\subsection{Translation solution}

Subsequently, we must also account for the motion of the droplet. We define the coordinate system such that the swimming velocity $\vec{U}$ is along the $x$ direction.
In the laboratory frame of reference, the boundary conditions of the Brinkman equations [Eq.~\ref{eq:BrinkmanEqnPolar}] are then given by $\left. u_r \right|_{r=R} = U \cos \theta$ and $\left. u_\theta \right|_{r=R} = -U \sin \theta$.
This gives the solution for a translating droplet,

    \begin{subequations}
    \label{eq:TranslationSolutionBrinkmanSquirmer}
    \begin{align}
        \psi &= U \left[ \frac{R^2 K_2\left(\frac{R}{h }\right)-2 h  r K_1\left(\frac{r}{h }\right)}{r K_0\left(\frac{R}{h }\right)} \right] \sin (\theta ),
        \\
        u_r &=  U \left[ \frac{R^2 K_2\left(\frac{R}{h }\right)-2 h  r K_1\left(\frac{r}{h }\right)}{r^2 K_0\left(\frac{R}{h }\right)} \right] \cos (\theta ),
        \\
        u_\theta &= U \left[ \frac{R^2 K_2\left(\frac{R}{h }\right)
        -2 h  r K_1\left(\frac{r}{h }\right)
        -2 r^2 K_0\left(\frac{r}{h }\right)
        }{r^2 K_0\left(\frac{R}{h }\right)} \right] \sin (\theta ),
        \\
        p &= U \left[ \frac{\mu R^2 }{h ^2 r } 
        \frac{K_2\left(\frac{R}{h }\right)}{K_0\left(\frac{R}{h }\right)}
        \right] \cos (\theta ).
    \end{align}
    \end{subequations}
To obtain the flow in the reference frame co-moving with the droplet, we add $-U \cos \theta$ to $u_r$, and $U \sin \theta$ to $u_\theta$. Evaluated at $r=R$ this gives $\vec{u}=0$, and $\vec{u} = - \vec{U}$ as $r \to \infty$. One may therefore also interpret this solution as a uniform flow moving around an obstacle.

\subsection{Swimming solution}

The complete solution for a swimming squirmer is obtained by combining the solutions for the surface actuation [$\psi_\text{act}$; Eq.~\ref{eq:PumpingStreamfunction}] and the translation [$\psi_\text{tr}$; Eq.~\ref{eq:TranslationSolutionBrinkmanSquirmer}], so the combined stream function for a squirmer in a Brinkman medium is
    \begin{equation}
    \label{eq:SquirmerStreamFunction}
        \psi_\text{swim} = \psi_\text{act} + \psi_\text{tr} - \phi_\text{rest} U r \sin \theta,
    \end{equation}
where $\phi_\text{rest}=0$ in the laboratory frame and $\phi_\text{rest}=1$ in the co-moving frame of reference. As before, the flow velocities are found using Eq.~\ref{eq:streamfunctionDerivatives}.

If the droplet is freely swimming, the swimming speed $U$ and the first mode $B_1$ are coupled to each other because the propulsion force is equal and opposite to the hydrodynamic drag.
For a squirmer in an unconfined 3D Stokesian fluid, this relationship is $B_1 = \frac{3}{2}U$.
To derive this relation in a Brinkman medium, one imposes that the total force on the droplet must vanish,

    \begin{align}
    \label{eq:ForceRelation}
    \vec{F}_\text{tot} = \oint_S \mathbf{\sigma} \cdot d\vec{S} = 0,
    \end{align}
where the hydrodynamics stress tensor is $\mathbf{\sigma} = -p\mathbf{I}+2\mu \mathbf{E}$, where $\mathbf{I}$ is the identity, the rate of strain tensor is $\mathbf{E} = (\vec{\nabla} \vec{u} + (\vec{\nabla} \vec{u})^T)/2$, and the velocity and pressure fields are combined from Eqs.~\ref{eq:PumpingSolutionBrinkmanSquirmer} and \ref{eq:TranslationSolutionBrinkmanSquirmer}.
Hence, one finds the relationship

    \begin{align}
    \label{eq:BrinkmanSpeedModeRelation}
    B_1 &= U \left(2+ \frac{R}{2h} \frac{K_0\left(\frac{R}{h }\right)}{K_1\left(\frac{R}{h }\right)} \right).
    \end{align}
This expression depends on the channel height, since $H = h \sqrt{12}$, as seen in Eq.~\ref{eq:BrinkmanEqn}.
In the limit of weak confinement, we have $U\to\frac{1}{2} B_1$ as $H\to\infty$, in agreement with the result for a squirmer in a 2D Stokes fluid (see e.g. \cite{gilpin2017vortex}). 
In the opposite limit, however, the velocity tends to zero as the channel height decreases, as expected.

In our experiments, we measure $U \sim \SI{22}{\micro\metre\per\second}$ and $B_1 \sim \SI{29.8}{\micro\metre\per\second}$, which gives $U/B_1\sim 0.74$. 
Thus, the droplet moves faster than the prediction of Eq.~\ref{eq:BrinkmanSpeedModeRelation}.
Indeed, the force-free condition is not necessarily true in confinement because the droplet can benefit from additional traction with the walls, in agreement with \cite{liu2016bimetallic}.
The speed-up may also be related to the spatial confinement of the chemical fields, which has a non-linear relation with the activity \cite{maass2016_swimming}. 
Therefore, in all our simulations, we do not enforce Eq.~\ref{eq:ForceRelation} but set the parameter values for $U$, $B_1$ and $B_2$ equal to the measured values directly [see Fig.~\ref{fig1}D].
This leads to an accurate model for the flows generated by the self-propelled droplets in confinement [Fig.~\ref{fig2}].

\section{Collective transport}
\label{sec:EntrainmentVolume}

\subsection{Simulation details}
\label{sec:CollectiveTransportSimulations}

We now consider collective entrainment by a school of self-propelled droplets [Fig.~\ref{fig4}A].
The swimmers all move in the same $\hat{\vec{x}}$ direction with speed $U$. 
The swimmer positions $\vec{r}_i$ are randomly distributed in the $xy$ plane with area fraction $\phi$, so the number density is $n=\phi/\pi R^2$.
In the dilute limit, the total flow generated by the $N$ droplets is given by the sum of the individual droplet flows [Eq.~\ref{eq:SquirmerStreamFunction}]. 
A tracer particle also experiences a short-ranged excluded volume repulsion to each droplet, $\vec{u}_\text{WCA}$, which we model with a Weeks-Chandler-Andersen (WCA) potential \cite{weeks1971role}.
Together with Brownian motion, the resulting equation of motion for the tracer position is

    \begin{align}
    \label{eq:ParticleEOM2}
    \frac{d \vec{r}_\text{T}}{dt} &= \sum_{i=1}^N \vec{u}_\text{tot}(\vec{r}_\text{T} - \vec{r}_i) + \sqrt{2D}\vec{\xi}(t),
    \end{align}
where $\vec{u}_\text{tot} = \vec{u}_\text{flow} + \vec{u}_\text{WCA}$.
A school of $N=500$ droplets is distributed over a rectangular area of length $L_x=1000\pi R$ and width $L_y=50R$, so the area fraction is $\phi=0.01$.
Note, we verified that using a larger simulation box did not change the results. 
We then integrate the equation of motion for $N_\text{T}=50$ tracer particles that are initially located at $x=0$ and spread over $-10R < y < 10R$.
The resulting dynamics are shown in Fig.~\ref{fig4}B and Video 3.

\subsection{Darwin drift}
\label{sec:StreamfunctionApproach}

Sir Charles Galton Darwin, grandson of the evolutionary biologist Charles Robert Darwin, demonstrated that the added mass of a body moving through an inviscid fluid is equal to the volume of liquid displaced by that body, multiplied with the fluid density \cite{darwin1953note, eames1994drift}. 
More recent insights also show that this ``Darwin drift'' volume is closely related to ``Stokes drift'' \cite{eames1999connection}.
Furthermore, the entrainment volume is important for fluid transport by micro-swimmers \cite{pushkin2012fluid}, swimming or flying forces may be estimated from hydrodynamics wakes \cite{dabiri2005estimation}, and fluid transport is intrinsically connected to biogenic mixing processes in the ocean \cite{katija2009viscosity, houghton2018vertically}.

To compute the total volume of liquid entrained by a swimming droplet, we follow the streamfunction approach used by Pushkin \textit{et al.} \cite{pushkin2012fluid}.
This method is illustrated in Fig.~\ref{fig4}C.
The streamlines that move around the swimmer body (in the co-moving frame of reference) are given by level sets of the streamfunction [Eq.~\ref{eq:SquirmerStreamFunction}].
This streamfunction can be written in 2D Cartesian coordinates as $\psi(x,y) = \int_A^B u_x dy - u_y dx$, which physically represents the volume flux of incompressible fluid crossing through any curve that connects the points $A$ and $B$.
Hence, the streamfunction of a uniform flow $u_x = -U$ is given by $\psi_0 = -Uy$. 
Far away from the self-propelled drop, the flow in the rest frame tends to this undisturbed flow, so $\psi \to \psi_0$ as $r\to\infty$.

We then consider a curtain of particles placed in front of the swimmer [see Fig.~\ref{fig4}C], initially located along the line from point ($x_\text{i}, 0$) to ($x_\text{i}, y_0$).
In the absence of a droplet, this curtain sweeps out a streamtube with rectangular area $\psi_0(x_\text{i}, y_0) \Delta t = (x_\text{i}-x_\text{f}) y_0$ during a time $\Delta t$.
With the droplet present, the curtain wraps around the drop [as in Fig.~\ref{fig3}A] and sweeps out an area equal to $\psi(x_\text{i}, y_0) \Delta t$.
The resulting entrainment area $A_\text{ent}$ is shaded blue in Fig.~\ref{fig4}C.
Next, we consider two more regions: 
First, the area enclosed by a stagnation streamline around the swimmer, denoted $A_* = A_\text{drop} + A_\text{wake}$, which comprises the droplet area $A_\text{drop} = \pi R^2$ [grey shade] and a wake area $A_\text{wake}$ [purple shade].
Second, the auxiliary area [green shade] is defined as

    \begin{equation}
    \label{eq:AuxiliaryArea}
        A_\text{aux} = 2 \int_{x_\text{f}}^{x_\text{i}} (y-y_0) dx,
    \end{equation}
where $y(x)$ describes the position of the streamline that passes through the points ($x_\text{i}, y_0$) and ($x_\text{f}, y_0$),
and the factor of 2 stems from the reflection in the $x$ axis.
Putting everything together, we have the relation

    \begin{equation}
        2 \psi_0 \Delta t + A_\text{aux} = 2 \psi \Delta t + A_\text{ent} + A_*.
    \end{equation}
Hence, in the limit of large $y_0$ and $|x_\text{i}|, |x_\text{f}|$ values, when $\psi \to \psi_0$, the entrainment volume $V_\text{ent} =  HA_\text{ent}$ is 

    \begin{align}
    \label{eq:EntrainmentArea}
        V_\text{ent} &= H(A_\text{aux} - A_*)
        \\
        &= H(A_\text{aux} - A_\text{drop} - A_\text{wake}),
    \end{align}
where $H=2R$ is the microfluidic channel height.

We first evaluate the auxiliary area, by finding and integrating $y(x)$.
This curve describes by the streamline that starts at the point $(x=\infty, y=y_0)$, so it is determined by the level set

    \begin{equation}
    \label{eq:LevelSet1}
    \psi(x,y(x)) = -U y_0,
    \end{equation}
which we aim to solve for $y(x)$. 
Inserting the streamfunction [Eq.~\ref{eq:SquirmerStreamFunction}] up to the second actuation mode ($n=2$), with $\phi_\text{rest}=1$, we have
$\psi_\text{act} + \psi_\text{tr} = U (y-y_0)$.
By defining $\epsilon= y-y_0$ and expanding $\psi$ to leading order in $\epsilon$, and subsequently noting that all terms of the form $K_n\left(\sqrt{x^2+y_0^2}/{h }\right)$ decay exponentially with $y_0$, then Eq.~\ref{eq:LevelSet1} can be simplified as

    \begin{align}
    \label{eq:LevelSet2}
    \epsilon U 
    &= 
    U \frac{R^2  y_0 K_2\left(\frac{R}{h }\right)}{\left(x^2+y_0^2\right) K_0\left(\frac{R}{h }\right)}
    -
    B_1 \frac{ h  R y_0 K_1\left(\frac{R}{h }\right)}{\left(x^2+y_0^2\right) K_0\left(\frac{R}{h }\right)}
    \nonumber \\
    &-
    b_2 \frac{2  h  R^2 x y_0 K_2\left(\frac{R}{h }\right)}{\left(x^2+y_0^2\right){}^2 K_1\left(\frac{R}{h }\right)}.
    \end{align}
This expression is then inserted into Eq.~\ref{eq:AuxiliaryArea}.
Integrating from $x_\text{f}=-\infty$ to $x_\text{i}=\infty$ gives

    \begin{align}
    \label{eq:AuxArea}
    A_\text{aux}
    &=
    2 \pi  R^2 \left(1 + \frac{\left(2 U-B_1\right) h K_1\left(\frac{R}{h }\right)}{U R K_0\left(\frac{R}{h }\right)}\right).
    \end{align}
Interestingly, this result does not depend on the dipole moment $b_2$, and also not on higher-order moments.
Moreover, this expression is finite for swimmers that exert a net force on the fluid (Eq.~\ref{eq:ForceRelation}) in this confined Brinkman medium.
This contrasts with an unconfined Stokesian fluid, where the Darwin drift volume diverges as the distance traveled becomes large \cite{eames2003fluid, shaik2020drag, chisholm2017drift, pushkin2012fluid, chisholm2018partial}.

We then shift our attention to the area $A_*$ enclosed by the stagnation streamline given by the expression $\psi=0$. 
This expression holds true at $r=R$, on the surface of the droplet.
There is no wake if the dipole coefficient $|\beta| \leq 1$, so $V_\text{wake} =0$ and $A_* = A_\text{drop}$.
For our droplets we measured $\beta \approx -0.3$. 
Indeed, we did not observe a wake in our flow field measurements [Fig.~\ref{fig2}].
Still, we will discuss non-zero wake sizes for the case of $|\beta| > 1$ in \S\ref{sec:LargeDipole}.


Finally, after combining the results [Eq.~\ref{eq:AuxiliaryArea}-\ref{eq:AuxArea}], the entrainment volume of a single droplet is given by 

    \begin{align}
    \label{eq:EntrainmentVolumeAnalytical}
    V_\text{ent}
    &= 
    \frac{4}{3}\pi R^3
    \left(
    \frac{3}{2}+\frac{3 h  \left(2 U - B_1\right) K_1\left(\frac{R}{h }\right)}{R U K_0\left(\frac{R}{h }\right)} \right ) - V_\text{wake}.
    \end{align}
Inserting the experimental values for our active droplets, we find $V_\text{ent}/ V_\text{drop} \approx 2.93$.    
Hence, the effective cargo capacity of a micro-carrier can be enlarged significantly by this hydrodynamic entrainment effect.
In general, for different micro-swimmers, it is also possible to feature a negative entrainment volume. 
This could be because they have a large wake volume that results from a strong dipole coefficient $|\beta|$ or higher-order flow structures. 
The first term in Eq.~\ref{eq:EntrainmentVolumeAnalytical} can also be negative for slowly moving swimmers, with $B_1\gg U$, which effectively pump fluid backwards.
It would be of great interest to measure $V_\text{ent}$ for different micro-organisms and link this collective transport to their biofunctionalities. 

\subsection{Numerical approach}
\label{subsec:NumericalApproach}

To verify this result [Eq.~\ref{eq:EntrainmentVolumeAnalytical}], the entrainment volume can be computed numerically by integrating the particle displacement [Eq.~\ref{eq:EntrainmentVolume2D}].
We can rewrite this expression as

    \begin{align}
    \label{eq:EntrainmentVolumeNum}
    V_\text{ent} 
    &= 2H\int_0^\infty  \Delta_x da
    \nonumber \\
    &= 4R^3\int_{-\infty}^\infty \frac{a \Delta_x}{R^2} d \text{log}\left( \frac{a}{R}\right).
    \end{align}    
Here the second integral form helps with taking care of the divergent entrainment directly in front of the droplet, when $\Delta_x \to \infty$ as $a \to 0$ but the integrand $I(a)=a\Delta_x/R^2$ remains finite.
We computed $\Delta_x(a)$ by numerically integrating the equation of motion [Eq.~\ref{eq:ParticleEOM}] with constant $\lambda=10^5 R$ and $b=\lambda/2$.
Using the values from our experiments, we again find $V_\text{ent}/V_\text{drop} \approx 2.93$, in agreement with the analytical result of Eq.~\ref{eq:EntrainmentVolumeAnalytical}.

\subsection{Collective entrainment velocity}
\label{subsec:CollectiveEntrainmentVelocity}
 
For a dilute school of droplets with area fraction $\phi \ll 1$, the mean entrainment velocity of the fluid in our quasi-2D geometry can be written as 
    \begin{equation}
    \label{eq:EntrainmentVel1}
    U_\text{ent} = n U A_\text{ent},
    \end{equation}  
where the first two factors encode the encounter rate, and $V_\text{ent} = H A_\text{ent}$ is the Darwin drift volume due a single encounter computed in Eq.~\ref{eq:EntrainmentVolumeAnalytical} above.
Thus, the mean entrainment velocity is

    \begin{align}
    \label{eq:TransportVelocity2}
    \frac{ U_\text{ent}}{U}
    &=  n A_\text{ent}
    =
    \frac{\phi V_\text{ent}}{\pi R^2 H}
    \\
    &= \phi  \left(1 + \frac{2 \left(2 U-B_1\right) h K_1\left(\frac{R}{h }\right)}{U R K_0\left(\frac{R}{h }\right)}\right)
    -\frac{\phi V_\text{wake}}{\pi R^2 H}. \nonumber
     \end{align}   
Inserting $\phi=0.01$ and our experimental values gives $U_\text{ent} \approx 1.95\,\phi U \approx \SI{0.43}{\micro\metre\per\second}$, which agrees with the simulations shown in Fig.~\ref{fig4}B.

\subsection{Large dipole coefficients}
\label{sec:LargeDipole}

\begin{figure}[t]
    \includegraphics[width=.5\linewidth]{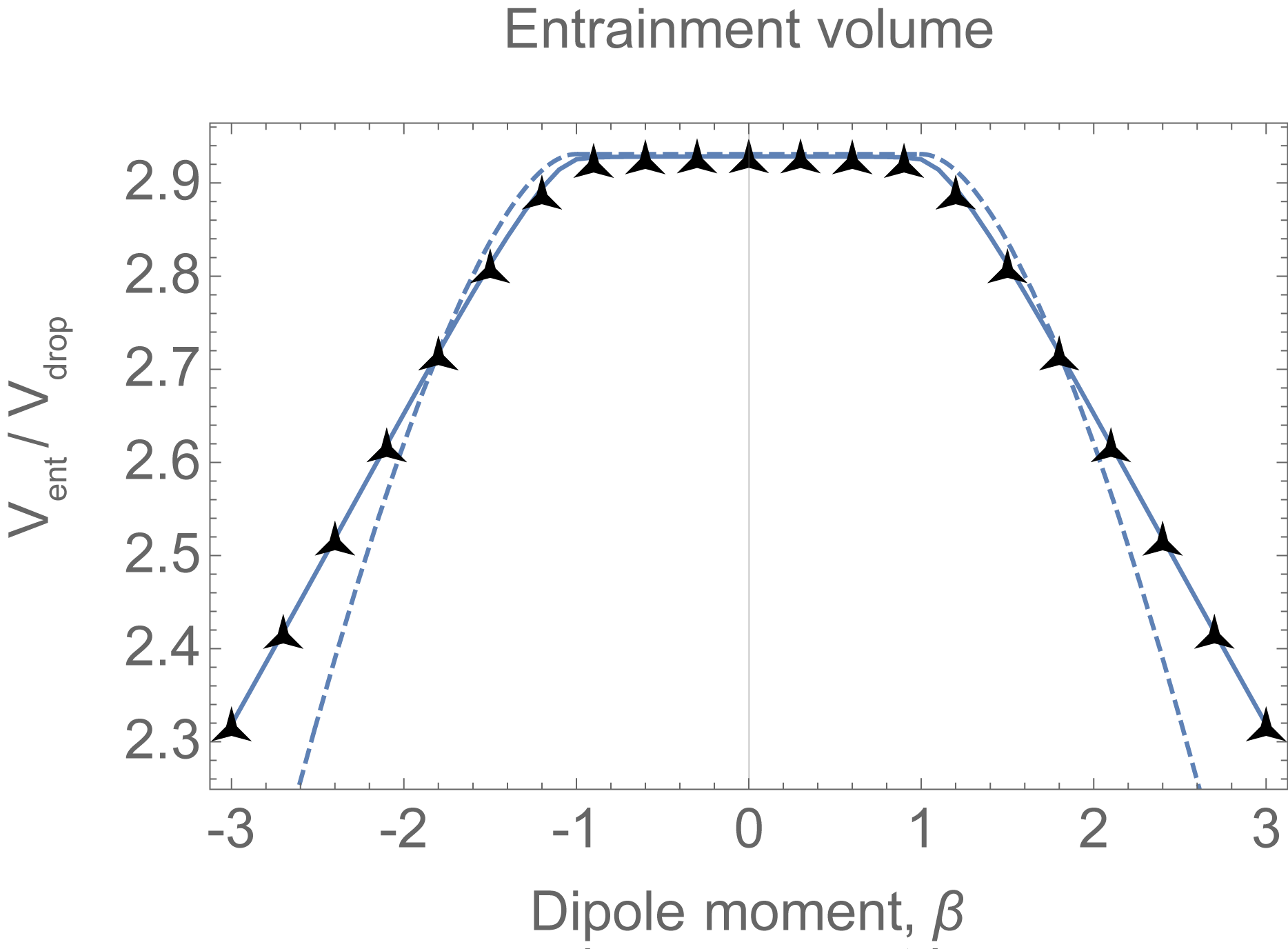}
    \caption{
    Entrainment volume of a squirmer in a confined Brinkman fluid as a function of the dipole coefficient.
    The black markers show the result obtained from integrating Eq.~\ref{eq:EntrainmentVolumeNum} numerically.
    The solid blue line shows the result obtained from integrating Eqs.~\ref{eq:AuxiliaryArea} and \ref{eq:WakeArea} numerically.
    The dashed blue line shows the analytical approximation of Eqs.~\ref{eq:EntrainmentVolumeAnalytical} and \ref{eq:WakeVolumeAnalytical}.
    }
    \label{figS1}
\end{figure}

For our active droplets, we measured a dipole coefficient of $\beta \approx -0.3$ and we did not observe a wake volume.
However, other types of micro-swimmers might have a larger dipole coefficient and a finite wake volume $V_\text{wake}$.
Therefore, for completeness, we describe how the model can be extended for $|\beta|>1$.
In that case, the streamline $\psi=0$ describes a separatrix that encloses both the droplet and its wake [Fig.~\ref{fig4}C; purple]. 

Pushkin \textit{et al.} \cite{pushkin2012fluid} derived this wake volume of a squirmer in an unconfined Stokesian fluid to be

    \begin{align}
    \label{eq:WakeVolumeAnalytical}
    \frac{V_\text{wake}}{V_\text{drop}}
    &= 
    \frac{r_*^3-1}{2} 
    \\
    &+ \frac{1}{\beta} \left( \frac{31}{30}+\log \left(\frac{r_{*}+1}{2}\right)-r_{*}+\frac{r_{*}^{2}}{2}-\frac{r_{*}^{3}}{3}-\frac{r_{*}^{5}}{5}\right), \nonumber
    \end{align}
where $r_*(\beta)$ satisfies the expression

    \begin{align}
    \label{eq:WakeVolumeAnalytical2}
    r_{*}\left(r_{*}^{2}+r_{*}+1\right)-(3 \beta / 2)\left(r_{*}+1\right)=0.
    \end{align}
The wake volume in the Brinkman fluid is approximately equal to the wake volume in the Stokes fluid, to first approximation, because the squirming boundary condition on the surface of the micro-swimmer [Eq.~\ref{eq:SquirmerModes1}] is exactly the same. 

To demonstrate this, the size of the wake can also be obtained numerically. The area $A_*$ is given by the separatrix streamline $\psi=0$ that encloses the drop and the wake. Analogous to Eq.~\ref{eq:LevelSet1}, we can solve for the separatrix position $y_\text{sep}(x)$ that satisfies
    \begin{equation}
    \label{eq:Separatrix}
        \psi(x,y_\text{sep}(x))=0.
    \end{equation}
The wake size is then given by
    \begin{equation}
    \label{eq:WakeArea}
        A_\text{wake} = 2 \int_{-\infty}^{\infty} y_\text{sep}(x) dx - A_\text{drop}.
    \end{equation}
This reduces to $A_\text{wake} = 0$ in the case of $|\beta| \leq 1$, but it is finite otherwise. 

Fig.~\ref{figS1} shows the entrainment volume of a squirmer in a confined Brinkman fluid as a function of the dipole coefficient $\beta$.
The black markers show the result obtained from integrating Eq.~\ref{eq:EntrainmentVolumeNum} numerically.
The solid blue line shows the result obtained from integrating Eqs.~\ref{eq:AuxiliaryArea} and \ref{eq:WakeArea} numerically.
The dashed blue line shows the analytical approximation of Eqs.~\ref{eq:EntrainmentVolumeAnalytical} and \ref{eq:WakeVolumeAnalytical}.
As before, we use the measured parameters for our active drops, $R \sim \SI{25}{\micro\metre}$, $H \sim \SI{50}{\micro\metre}$, $U \sim \SI{22}{\micro\metre\per\second}$, $B_1 \sim \SI{30}{\micro\metre\per\second}$, and we varied the second moment $B_2 = \beta B_1$.
As expected, the entrainment volume is constant for $|\beta| \leq 1$ and smaller for the cases of $|\beta|>1$.
Both the auxiliary volume and the wake volume are invariant under $\beta \to -\beta$, so the entrainment volume is the same for pushers and pullers. Finally, the enhanced cargo transport is robust up to large dipole coefficients.

\section{Video captions}
\label{sec:VideoCaptions}
The respective video files have been deposited on arXiv as ancillary data.

\textbf{Video 1} --- Experiment showing particle displacement due to an active droplet, which generates flows as it swims along a straight line. The droplet is confined in a microfluidic chamber of height equal to the droplet diameter. Particle trajectories are highlighted in different colours.

\textbf{Video 2} --- Simulation of particle entrainment by a confined self-propelled droplet, in the absence of Brownian motion, using the model for a squirmer in a Brinkman fluid. Shown is the deformation of an initially uniform rectangular grid of tracer particles, after a droplet swims from $x/R=-5$ to $5$. Coloured trajectories (from violet to red) show the motion of individual particles.

\textbf{Video 3} --- Entrainment by a school of micro-swimmers. The self-propelled droplets (black) move collectively along the positive $x$ direction, transporting particles (colours) that initially started at $x=0$. Brownian motion is included. Averaged over time, the particles are transported with the mean entrainment velocity $\vec{U}_\text{ent}$ predicted by Eq.~\ref{eq:TransportVelocity}.

%% file: main.bbl
\begin{thebibliography}{95}%
\makeatletter
\providecommand \@ifxundefined [1]{%
 \@ifx{#1\undefined}
}%
\providecommand \@ifnum [1]{%
 \ifnum #1\expandafter \@firstoftwo
 \else \expandafter \@secondoftwo
 \fi
}%
\providecommand \@ifx [1]{%
 \ifx #1\expandafter \@firstoftwo
 \else \expandafter \@secondoftwo
 \fi
}%
\providecommand \natexlab [1]{#1}%
\providecommand \enquote  [1]{``#1''}%
\providecommand \bibnamefont  [1]{#1}%
\providecommand \bibfnamefont [1]{#1}%
\providecommand \citenamefont [1]{#1}%
\providecommand \href@noop [0]{\@secondoftwo}%
\providecommand \href [0]{\begingroup \@sanitize@url \@href}%
\providecommand \@href[1]{\@@startlink{#1}\@@href}%
\providecommand \@@href[1]{\endgroup#1\@@endlink}%
\providecommand \@sanitize@url [0]{\catcode `\\12\catcode `\$12\catcode
  `\&12\catcode `\#12\catcode `\^12\catcode `\_12\catcode `\%12\relax}%
\providecommand \@@startlink[1]{}%
\providecommand \@@endlink[0]{}%
\providecommand \url  [0]{\begingroup\@sanitize@url \@url }%
\providecommand \@url [1]{\endgroup\@href {#1}{\urlprefix }}%
\providecommand \urlprefix  [0]{URL }%
\providecommand \Eprint [0]{\href }%
\providecommand \doibase [0]{http://dx.doi.org/}%
\providecommand \selectlanguage [0]{\@gobble}%
\providecommand \bibinfo  [0]{\@secondoftwo}%
\providecommand \bibfield  [0]{\@secondoftwo}%
\providecommand \translation [1]{[#1]}%
\providecommand \BibitemOpen [0]{}%
\providecommand \bibitemStop [0]{}%
\providecommand \bibitemNoStop [0]{.\EOS\space}%
\providecommand \EOS [0]{\spacefactor3000\relax}%
\providecommand \BibitemShut  [1]{\csname bibitem#1\endcsname}%
\let\auto@bib@innerbib\@empty
\bibitem [{\citenamefont {Purcell}(1977)}]{purcell1977life}%
  \BibitemOpen
  \bibfield  {author} {\bibinfo {author} {\bibfnamefont {E.~M.}\ \bibnamefont
  {Purcell}},\ }\href {\doibase 10.1119/1.10903} {\bibfield  {journal}
  {\bibinfo  {journal} {American J. Phys.}\ }\textbf {\bibinfo {volume} {45}},\
  \bibinfo {pages} {3} (\bibinfo {year} {1977})}\BibitemShut {NoStop}%
\bibitem [{\citenamefont {Darwin}(1953)}]{darwin1953note}%
  \BibitemOpen
  \bibfield  {author} {\bibinfo {author} {\bibfnamefont {C.~G.}\ \bibnamefont
  {Darwin}},\ }in\ \href {\doibase 10.1017/S0305004100028449} {\emph {\bibinfo
  {booktitle} {Math. Proc. Cambridge Phil. Soc.}}},\ Vol.~\bibinfo {volume}
  {49}\ (\bibinfo {organization} {Cambridge University Press},\ \bibinfo {year}
  {1953})\ pp.\ \bibinfo {pages} {342--354}\BibitemShut {NoStop}%
\bibitem [{\citenamefont {Eames}\ \emph {et~al.}(1994)\citenamefont {Eames},
  \citenamefont {Belcher},\ and\ \citenamefont {Hunt}}]{eames1994drift}%
  \BibitemOpen
  \bibfield  {author} {\bibinfo {author} {\bibfnamefont {I.}~\bibnamefont
  {Eames}}, \bibinfo {author} {\bibfnamefont {S.~E.}\ \bibnamefont {Belcher}},
  \ and\ \bibinfo {author} {\bibfnamefont {J.~C.~R.}\ \bibnamefont {Hunt}},\
  }\href {\doibase 10.1017/S0022112094002338} {\bibfield  {journal} {\bibinfo
  {journal} {J. Fluid Mech.}\ }\textbf {\bibinfo {volume} {275}},\ \bibinfo
  {pages} {201} (\bibinfo {year} {1994})}\BibitemShut {NoStop}%
\bibitem [{\citenamefont {Eames}\ \emph {et~al.}(2003)\citenamefont {Eames},
  \citenamefont {Gobby},\ and\ \citenamefont {Dalziel}}]{eames2003fluid}%
  \BibitemOpen
  \bibfield  {author} {\bibinfo {author} {\bibfnamefont {I.}~\bibnamefont
  {Eames}}, \bibinfo {author} {\bibfnamefont {D.}~\bibnamefont {Gobby}}, \ and\
  \bibinfo {author} {\bibfnamefont {S.}~\bibnamefont {Dalziel}},\ }\href
  {\doibase 10.1017/S0022112003004361} {\bibfield  {journal} {\bibinfo
  {journal} {J. Fluid Mech.}\ }\textbf {\bibinfo {volume} {485}},\ \bibinfo
  {pages} {67} (\bibinfo {year} {2003})}\BibitemShut {NoStop}%
\bibitem [{\citenamefont {Shaik}\ and\ \citenamefont
  {Ardekani}(2020)}]{shaik2020drag}%
  \BibitemOpen
  \bibfield  {author} {\bibinfo {author} {\bibfnamefont {V.~A.}\ \bibnamefont
  {Shaik}}\ and\ \bibinfo {author} {\bibfnamefont {A.~M.}\ \bibnamefont
  {Ardekani}},\ }\href {\doibase 10.1103/PhysRevFluids.5.013604} {\bibfield
  {journal} {\bibinfo  {journal} {Phys. Rev. Fluids}\ }\textbf {\bibinfo
  {volume} {5}},\ \bibinfo {pages} {013604} (\bibinfo {year}
  {2020})}\BibitemShut {NoStop}%
\bibitem [{\citenamefont {Chisholm}\ and\ \citenamefont
  {Khair}(2017)}]{chisholm2017drift}%
  \BibitemOpen
  \bibfield  {author} {\bibinfo {author} {\bibfnamefont {N.~G.}\ \bibnamefont
  {Chisholm}}\ and\ \bibinfo {author} {\bibfnamefont {A.~S.}\ \bibnamefont
  {Khair}},\ }\href {\doibase 10.1103/PhysRevFluids.2.064101} {\bibfield
  {journal} {\bibinfo  {journal} {Phys. Rev. Fluids}\ }\textbf {\bibinfo
  {volume} {2}},\ \bibinfo {pages} {064101} (\bibinfo {year}
  {2017})}\BibitemShut {NoStop}%
\bibitem [{\citenamefont {Pushkin}\ \emph {et~al.}(2013)\citenamefont
  {Pushkin}, \citenamefont {Shum},\ and\ \citenamefont
  {Yeomans}}]{pushkin2012fluid}%
  \BibitemOpen
  \bibfield  {author} {\bibinfo {author} {\bibfnamefont {D.~O.}\ \bibnamefont
  {Pushkin}}, \bibinfo {author} {\bibfnamefont {H.}~\bibnamefont {Shum}}, \
  and\ \bibinfo {author} {\bibfnamefont {J.~M.}\ \bibnamefont {Yeomans}},\
  }\href {\doibase 10.1017/jfm.2013.208} {\bibfield  {journal} {\bibinfo
  {journal} {J. Fluid Mech.}\ }\textbf {\bibinfo {volume} {726}},\ \bibinfo
  {pages} {5} (\bibinfo {year} {2013})}\BibitemShut {NoStop}%
\bibitem [{\citenamefont {Chisholm}\ and\ \citenamefont
  {Khair}(2018)}]{chisholm2018partial}%
  \BibitemOpen
  \bibfield  {author} {\bibinfo {author} {\bibfnamefont {N.~G.}\ \bibnamefont
  {Chisholm}}\ and\ \bibinfo {author} {\bibfnamefont {A.~S.}\ \bibnamefont
  {Khair}},\ }\href {\doibase 10.1103/PhysRevFluids.3.014501} {\bibfield
  {journal} {\bibinfo  {journal} {Phys. Rev. Fluids}\ }\textbf {\bibinfo
  {volume} {3}},\ \bibinfo {pages} {014501} (\bibinfo {year}
  {2018})}\BibitemShut {NoStop}%
\bibitem [{\citenamefont {Wu}\ and\ \citenamefont
  {Libchaber}(2000)}]{wu2000particle}%
  \BibitemOpen
  \bibfield  {author} {\bibinfo {author} {\bibfnamefont {X.-L.}\ \bibnamefont
  {Wu}}\ and\ \bibinfo {author} {\bibfnamefont {A.}~\bibnamefont {Libchaber}},\
  }\href {\doibase 10.1103/physrevlett.84.3017} {\bibfield  {journal} {\bibinfo
   {journal} {Phys. Rev. Lett.}\ }\textbf {\bibinfo {volume} {84}},\ \bibinfo
  {pages} {3017} (\bibinfo {year} {2000})}\BibitemShut {NoStop}%
\bibitem [{\citenamefont {Kim}\ and\ \citenamefont
  {Breuer}(2004)}]{kim2004enhanced}%
  \BibitemOpen
  \bibfield  {author} {\bibinfo {author} {\bibfnamefont {M.~J.}\ \bibnamefont
  {Kim}}\ and\ \bibinfo {author} {\bibfnamefont {K.~S.}\ \bibnamefont
  {Breuer}},\ }\href {\doibase 10.1063/1.1787527} {\bibfield  {journal}
  {\bibinfo  {journal} {Phys. Fluids}\ }\textbf {\bibinfo {volume} {16}},\
  \bibinfo {pages} {L78} (\bibinfo {year} {2004})}\BibitemShut {NoStop}%
\bibitem [{\citenamefont {Leptos}\ \emph {et~al.}(2009)\citenamefont {Leptos},
  \citenamefont {Guasto}, \citenamefont {Gollub}, \citenamefont {Pesci},\ and\
  \citenamefont {Goldstein}}]{leptos2009_dynamics}%
  \BibitemOpen
  \bibfield  {author} {\bibinfo {author} {\bibfnamefont {K.~C.}\ \bibnamefont
  {Leptos}}, \bibinfo {author} {\bibfnamefont {J.~S.}\ \bibnamefont {Guasto}},
  \bibinfo {author} {\bibfnamefont {J.~P.}\ \bibnamefont {Gollub}}, \bibinfo
  {author} {\bibfnamefont {A.~I.}\ \bibnamefont {Pesci}}, \ and\ \bibinfo
  {author} {\bibfnamefont {R.~E.}\ \bibnamefont {Goldstein}},\ }\href {\doibase
  10.1103/PhysRevLett.103.198103} {\bibfield  {journal} {\bibinfo  {journal}
  {Phys. Rev. Lett.}\ }\textbf {\bibinfo {volume} {103}},\ \bibinfo {pages}
  {198103} (\bibinfo {year} {2009})}\BibitemShut {NoStop}%
\bibitem [{\citenamefont {Thiffeault}\ and\ \citenamefont
  {Childress}(2010)}]{thiffeault2010stirring}%
  \BibitemOpen
  \bibfield  {author} {\bibinfo {author} {\bibfnamefont {J.-L.}\ \bibnamefont
  {Thiffeault}}\ and\ \bibinfo {author} {\bibfnamefont {S.}~\bibnamefont
  {Childress}},\ }\href {\doibase 10.1016/j.physleta.2010.06.043} {\bibfield
  {journal} {\bibinfo  {journal} {Phys. Lett. A}\ }\textbf {\bibinfo {volume}
  {374}},\ \bibinfo {pages} {3487} (\bibinfo {year} {2010})}\BibitemShut
  {NoStop}%
\bibitem [{\citenamefont {Lin}\ \emph {et~al.}(2011)\citenamefont {Lin},
  \citenamefont {Thiffeault},\ and\ \citenamefont
  {Childress}}]{lin2011_stirring}%
  \BibitemOpen
  \bibfield  {author} {\bibinfo {author} {\bibfnamefont {Z.}~\bibnamefont
  {Lin}}, \bibinfo {author} {\bibfnamefont {J.-L.}\ \bibnamefont {Thiffeault}},
  \ and\ \bibinfo {author} {\bibfnamefont {S.}~\bibnamefont {Childress}},\
  }\href {\doibase 10.1017/s002211201000563x} {\bibfield  {journal} {\bibinfo
  {journal} {J. Fluid Mech.}\ }\textbf {\bibinfo {volume} {669}},\ \bibinfo
  {pages} {167} (\bibinfo {year} {2011})}\BibitemShut {NoStop}%
\bibitem [{\citenamefont {Pushkin}\ and\ \citenamefont
  {Yeomans}(2013)}]{pushkin2013fluid}%
  \BibitemOpen
  \bibfield  {author} {\bibinfo {author} {\bibfnamefont {D.~O.}\ \bibnamefont
  {Pushkin}}\ and\ \bibinfo {author} {\bibfnamefont {J.~M.}\ \bibnamefont
  {Yeomans}},\ }\href {\doibase 10.1103/PhysRevLett.111.188101} {\bibfield
  {journal} {\bibinfo  {journal} {Phys. Rev. Lett.}\ }\textbf {\bibinfo
  {volume} {111}},\ \bibinfo {pages} {188101} (\bibinfo {year}
  {2013})}\BibitemShut {NoStop}%
\bibitem [{\citenamefont {Morozov}\ and\ \citenamefont
  {Marenduzzo}(2014)}]{morozov2014enhanced}%
  \BibitemOpen
  \bibfield  {author} {\bibinfo {author} {\bibfnamefont {A.}~\bibnamefont
  {Morozov}}\ and\ \bibinfo {author} {\bibfnamefont {D.}~\bibnamefont
  {Marenduzzo}},\ }\href {\doibase 10.1039/C3SM52201F} {\bibfield  {journal}
  {\bibinfo  {journal} {Soft Matter}\ }\textbf {\bibinfo {volume} {10}},\
  \bibinfo {pages} {2748} (\bibinfo {year} {2014})}\BibitemShut {NoStop}%
\bibitem [{\citenamefont {Peng}\ \emph {et~al.}(2016)\citenamefont {Peng},
  \citenamefont {Lai}, \citenamefont {Tai}, \citenamefont {Zhang},
  \citenamefont {Xu},\ and\ \citenamefont {Cheng}}]{peng2016diffusion}%
  \BibitemOpen
  \bibfield  {author} {\bibinfo {author} {\bibfnamefont {Y.}~\bibnamefont
  {Peng}}, \bibinfo {author} {\bibfnamefont {L.}~\bibnamefont {Lai}}, \bibinfo
  {author} {\bibfnamefont {Y.-S.}\ \bibnamefont {Tai}}, \bibinfo {author}
  {\bibfnamefont {K.}~\bibnamefont {Zhang}}, \bibinfo {author} {\bibfnamefont
  {X.}~\bibnamefont {Xu}}, \ and\ \bibinfo {author} {\bibfnamefont
  {X.}~\bibnamefont {Cheng}},\ }\href {\doibase 10.1103/PhysRevLett.116.068303}
  {\bibfield  {journal} {\bibinfo  {journal} {Phys. Rev. Lett.}\ }\textbf
  {\bibinfo {volume} {116}},\ \bibinfo {pages} {068303} (\bibinfo {year}
  {2016})}\BibitemShut {NoStop}%
\bibitem [{\citenamefont {Jeanneret}\ \emph {et~al.}(2016)\citenamefont
  {Jeanneret}, \citenamefont {Pushkin}, \citenamefont {Kantsler},\ and\
  \citenamefont {Polin}}]{jeanneret2016_entrainment}%
  \BibitemOpen
  \bibfield  {author} {\bibinfo {author} {\bibfnamefont {R.}~\bibnamefont
  {Jeanneret}}, \bibinfo {author} {\bibfnamefont {D.~O.}\ \bibnamefont
  {Pushkin}}, \bibinfo {author} {\bibfnamefont {V.}~\bibnamefont {Kantsler}}, \
  and\ \bibinfo {author} {\bibfnamefont {M.}~\bibnamefont {Polin}},\ }\href
  {\doibase 10.1038/ncomms12518} {\bibfield  {journal} {\bibinfo  {journal}
  {Nat. Commun.}\ }\textbf {\bibinfo {volume} {7}},\ \bibinfo {pages} {12518}
  (\bibinfo {year} {2016})}\BibitemShut {NoStop}%
\bibitem [{\citenamefont {Guzm\'an-Lastra}\ \emph {et~al.}(2021)\citenamefont
  {Guzm\'an-Lastra}, \citenamefont {L\"owen},\ and\ \citenamefont
  {Mathijssen}}]{mathijssen2021active}%
  \BibitemOpen
  \bibfield  {author} {\bibinfo {author} {\bibfnamefont {F.}~\bibnamefont
  {Guzm\'an-Lastra}}, \bibinfo {author} {\bibfnamefont {H.}~\bibnamefont
  {L\"owen}}, \ and\ \bibinfo {author} {\bibfnamefont {A.~J. T.~M.}\
  \bibnamefont {Mathijssen}},\ }\href {\doibase 10.1038/s41467-021-22029-y}
  {\bibfield  {journal} {\bibinfo  {journal} {Nat. Commun.}\ }\textbf {\bibinfo
  {volume} {12}},\ \bibinfo {pages} {1906} (\bibinfo {year}
  {2021})}\BibitemShut {NoStop}%
\bibitem [{\citenamefont {Katija}\ and\ \citenamefont
  {Dabiri}(2009)}]{katija2009viscosity}%
  \BibitemOpen
  \bibfield  {author} {\bibinfo {author} {\bibfnamefont {K.}~\bibnamefont
  {Katija}}\ and\ \bibinfo {author} {\bibfnamefont {J.~O.}\ \bibnamefont
  {Dabiri}},\ }\href {\doibase 10.1038/nature08207} {\bibfield  {journal}
  {\bibinfo  {journal} {Nature}\ }\textbf {\bibinfo {volume} {460}},\ \bibinfo
  {pages} {624} (\bibinfo {year} {2009})}\BibitemShut {NoStop}%
\bibitem [{\citenamefont {Subramanian}(2010)}]{subramanian2010viscosity}%
  \BibitemOpen
  \bibfield  {author} {\bibinfo {author} {\bibfnamefont {G.}~\bibnamefont
  {Subramanian}},\ }\href {https://www.jstor.org/stable/24111768} {\bibfield
  {journal} {\bibinfo  {journal} {Curr. Sci.}\ }\textbf {\bibinfo {volume}
  {98}},\ \bibinfo {pages} {1103} (\bibinfo {year} {2010})}\BibitemShut
  {NoStop}%
\bibitem [{\citenamefont {Nawroth}\ and\ \citenamefont
  {Dabiri}(2014)}]{nawroth2014induced}%
  \BibitemOpen
  \bibfield  {author} {\bibinfo {author} {\bibfnamefont {J.~C.}\ \bibnamefont
  {Nawroth}}\ and\ \bibinfo {author} {\bibfnamefont {J.~O.}\ \bibnamefont
  {Dabiri}},\ }\href {\doibase 10.1063/1.4893537} {\bibfield  {journal}
  {\bibinfo  {journal} {Phys. Fluids}\ }\textbf {\bibinfo {volume} {26}},\
  \bibinfo {pages} {091108} (\bibinfo {year} {2014})}\BibitemShut {NoStop}%
\bibitem [{\citenamefont {Houghton}\ \emph {et~al.}(2018)\citenamefont
  {Houghton}, \citenamefont {Koseff}, \citenamefont {Monismith},\ and\
  \citenamefont {Dabiri}}]{houghton2018vertically}%
  \BibitemOpen
  \bibfield  {author} {\bibinfo {author} {\bibfnamefont {I.~A.}\ \bibnamefont
  {Houghton}}, \bibinfo {author} {\bibfnamefont {J.~R.}\ \bibnamefont
  {Koseff}}, \bibinfo {author} {\bibfnamefont {S.~G.}\ \bibnamefont
  {Monismith}}, \ and\ \bibinfo {author} {\bibfnamefont {J.~O.}\ \bibnamefont
  {Dabiri}},\ }\href {\doibase 10.1038/s41586-018-0044-z} {\bibfield  {journal}
  {\bibinfo  {journal} {Nature}\ }\textbf {\bibinfo {volume} {556}},\ \bibinfo
  {pages} {497} (\bibinfo {year} {2018})}\BibitemShut {NoStop}%
\bibitem [{\citenamefont {Mathijssen}\ \emph {et~al.}(2019)\citenamefont
  {Mathijssen}, \citenamefont {Culver}, \citenamefont {Bhamla},\ and\
  \citenamefont {Prakash}}]{mathijssen2019collective}%
  \BibitemOpen
  \bibfield  {author} {\bibinfo {author} {\bibfnamefont {A.~J.}\ \bibnamefont
  {Mathijssen}}, \bibinfo {author} {\bibfnamefont {J.}~\bibnamefont {Culver}},
  \bibinfo {author} {\bibfnamefont {M.~S.}\ \bibnamefont {Bhamla}}, \ and\
  \bibinfo {author} {\bibfnamefont {M.}~\bibnamefont {Prakash}},\ }\href
  {\doibase 10.1038/s41586-019-1387-9} {\bibfield  {journal} {\bibinfo
  {journal} {Nature}\ }\textbf {\bibinfo {volume} {571}},\ \bibinfo {pages}
  {560} (\bibinfo {year} {2019})}\BibitemShut {NoStop}%
\bibitem [{\citenamefont {Ortlieb}\ \emph {et~al.}(2019)\citenamefont
  {Ortlieb}, \citenamefont {Rafa\"{\i}}, \citenamefont {Peyla}, \citenamefont
  {Wagner},\ and\ \citenamefont {John}}]{ortlieb2019statistics}%
  \BibitemOpen
  \bibfield  {author} {\bibinfo {author} {\bibfnamefont {L.}~\bibnamefont
  {Ortlieb}}, \bibinfo {author} {\bibfnamefont {S.}~\bibnamefont {Rafa\"{\i}}},
  \bibinfo {author} {\bibfnamefont {P.}~\bibnamefont {Peyla}}, \bibinfo
  {author} {\bibfnamefont {C.}~\bibnamefont {Wagner}}, \ and\ \bibinfo {author}
  {\bibfnamefont {T.}~\bibnamefont {John}},\ }\href {\doibase
  10.1103/PhysRevLett.122.148101} {\bibfield  {journal} {\bibinfo  {journal}
  {Phys. Rev. Lett.}\ }\textbf {\bibinfo {volume} {122}},\ \bibinfo {pages}
  {148101} (\bibinfo {year} {2019})}\BibitemShut {NoStop}%
\bibitem [{\citenamefont {Magar}\ \emph {et~al.}(2003)\citenamefont {Magar},
  \citenamefont {Goto},\ and\ \citenamefont {Pedley}}]{magar2003nutrient}%
  \BibitemOpen
  \bibfield  {author} {\bibinfo {author} {\bibfnamefont {V.}~\bibnamefont
  {Magar}}, \bibinfo {author} {\bibfnamefont {T.}~\bibnamefont {Goto}}, \ and\
  \bibinfo {author} {\bibfnamefont {T.~J.}\ \bibnamefont {Pedley}},\ }\href
  {\doibase 10.1093/qjmam/56.1.65} {\bibfield  {journal} {\bibinfo  {journal}
  {Q. J. Mech. Appl. Math.}\ }\textbf {\bibinfo {volume} {56}},\ \bibinfo
  {pages} {65} (\bibinfo {year} {2003})}\BibitemShut {NoStop}%
\bibitem [{\citenamefont {Short}\ \emph {et~al.}(2006)\citenamefont {Short},
  \citenamefont {Solari}, \citenamefont {Ganguly}, \citenamefont {Powers},
  \citenamefont {Kessler},\ and\ \citenamefont {Goldstein}}]{short2006flows}%
  \BibitemOpen
  \bibfield  {author} {\bibinfo {author} {\bibfnamefont {M.~B.}\ \bibnamefont
  {Short}}, \bibinfo {author} {\bibfnamefont {C.~A.}\ \bibnamefont {Solari}},
  \bibinfo {author} {\bibfnamefont {S.}~\bibnamefont {Ganguly}}, \bibinfo
  {author} {\bibfnamefont {T.~R.}\ \bibnamefont {Powers}}, \bibinfo {author}
  {\bibfnamefont {J.~O.}\ \bibnamefont {Kessler}}, \ and\ \bibinfo {author}
  {\bibfnamefont {R.~E.}\ \bibnamefont {Goldstein}},\ }\href {\doibase
  10.1073/pnas.0600566103} {\bibfield  {journal} {\bibinfo  {journal} {Proc.
  Natl. Acad. Sci.}\ }\textbf {\bibinfo {volume} {103}},\ \bibinfo {pages}
  {8315} (\bibinfo {year} {2006})}\BibitemShut {NoStop}%
\bibitem [{\citenamefont {Michelin}\ and\ \citenamefont
  {Lauga}(2011)}]{michelin2011optimal}%
  \BibitemOpen
  \bibfield  {author} {\bibinfo {author} {\bibfnamefont {S.}~\bibnamefont
  {Michelin}}\ and\ \bibinfo {author} {\bibfnamefont {E.}~\bibnamefont
  {Lauga}},\ }\href {\doibase 10.1063/1.3642645} {\bibfield  {journal}
  {\bibinfo  {journal} {Phys. Fluids}\ }\textbf {\bibinfo {volume} {23}},\
  \bibinfo {pages} {101901} (\bibinfo {year} {2011})}\BibitemShut {NoStop}%
\bibitem [{\citenamefont {Tam}\ and\ \citenamefont
  {Hosoi}(2011)}]{tam2011optimal}%
  \BibitemOpen
  \bibfield  {author} {\bibinfo {author} {\bibfnamefont {D.}~\bibnamefont
  {Tam}}\ and\ \bibinfo {author} {\bibfnamefont {A.~E.}\ \bibnamefont
  {Hosoi}},\ }\href {\doibase 10.1073/pnas.1011185108} {\bibfield  {journal}
  {\bibinfo  {journal} {Proc. Natl. Acad. Sci.}\ }\textbf {\bibinfo {volume}
  {108}},\ \bibinfo {pages} {1001} (\bibinfo {year} {2011})}\BibitemShut
  {NoStop}%
\bibitem [{\citenamefont {Mathijssen}\ \emph
  {et~al.}(2018{\natexlab{a}})\citenamefont {Mathijssen}, \citenamefont
  {Guzm\'an-Lastra}, \citenamefont {Kaiser},\ and\ \citenamefont
  {L\"owen}}]{mathijssen2019nutrient}%
  \BibitemOpen
  \bibfield  {author} {\bibinfo {author} {\bibfnamefont {A.~J. T.~M.}\
  \bibnamefont {Mathijssen}}, \bibinfo {author} {\bibfnamefont
  {F.}~\bibnamefont {Guzm\'an-Lastra}}, \bibinfo {author} {\bibfnamefont
  {A.}~\bibnamefont {Kaiser}}, \ and\ \bibinfo {author} {\bibfnamefont
  {H.}~\bibnamefont {L\"owen}},\ }\href {\doibase
  10.1103/PhysRevLett.121.248101} {\bibfield  {journal} {\bibinfo  {journal}
  {Phys. Rev. Lett.}\ }\textbf {\bibinfo {volume} {121}},\ \bibinfo {pages}
  {248101} (\bibinfo {year} {2018}{\natexlab{a}})}\BibitemShut {NoStop}%
\bibitem [{\citenamefont {Papavassiliou}\ and\ \citenamefont
  {Alexander}(2015)}]{papavassiliou2015many}%
  \BibitemOpen
  \bibfield  {author} {\bibinfo {author} {\bibfnamefont {D.}~\bibnamefont
  {Papavassiliou}}\ and\ \bibinfo {author} {\bibfnamefont {G.~P.}\ \bibnamefont
  {Alexander}},\ }\href {\doibase 10.1209/0295-5075/110/44001} {\bibfield
  {journal} {\bibinfo  {journal} {Eur. Phys. Lett.}\ }\textbf {\bibinfo
  {volume} {110}},\ \bibinfo {pages} {44001} (\bibinfo {year}
  {2015})}\BibitemShut {NoStop}%
\bibitem [{\citenamefont {Shum}\ and\ \citenamefont
  {Yeomans}(2017)}]{shum2017entrainment}%
  \BibitemOpen
  \bibfield  {author} {\bibinfo {author} {\bibfnamefont {H.}~\bibnamefont
  {Shum}}\ and\ \bibinfo {author} {\bibfnamefont {J.~M.}\ \bibnamefont
  {Yeomans}},\ }\href {\doibase 10.1103/PhysRevFluids.2.113101} {\bibfield
  {journal} {\bibinfo  {journal} {Phys. Rev. Fluids}\ }\textbf {\bibinfo
  {volume} {2}},\ \bibinfo {pages} {113101} (\bibinfo {year}
  {2017})}\BibitemShut {NoStop}%
\bibitem [{\citenamefont {Mueller}\ and\ \citenamefont
  {Thiffeault}(2017)}]{mueller2017transport}%
  \BibitemOpen
  \bibfield  {author} {\bibinfo {author} {\bibfnamefont {P.}~\bibnamefont
  {Mueller}}\ and\ \bibinfo {author} {\bibfnamefont {J.-L.}\ \bibnamefont
  {Thiffeault}},\ }\href {\doibase 10.1103/PhysRevFluids.2.013103} {\bibfield
  {journal} {\bibinfo  {journal} {Phys. Rev. Fluids}\ }\textbf {\bibinfo
  {volume} {2}},\ \bibinfo {pages} {013103} (\bibinfo {year}
  {2017})}\BibitemShut {NoStop}%
\bibitem [{\citenamefont {Vaccari}\ \emph {et~al.}(2018)\citenamefont
  {Vaccari}, \citenamefont {Molaei}, \citenamefont {Leheny},\ and\
  \citenamefont {Stebe}}]{vaccari2018cargo}%
  \BibitemOpen
  \bibfield  {author} {\bibinfo {author} {\bibfnamefont {L.}~\bibnamefont
  {Vaccari}}, \bibinfo {author} {\bibfnamefont {M.}~\bibnamefont {Molaei}},
  \bibinfo {author} {\bibfnamefont {R.~L.}\ \bibnamefont {Leheny}}, \ and\
  \bibinfo {author} {\bibfnamefont {K.~J.}\ \bibnamefont {Stebe}},\ }\href
  {\doibase 10.1039/C8SM00481A} {\bibfield  {journal} {\bibinfo  {journal}
  {Soft Matter}\ }\textbf {\bibinfo {volume} {14}},\ \bibinfo {pages} {5643}
  (\bibinfo {year} {2018})}\BibitemShut {NoStop}%
\bibitem [{\citenamefont {Purushothaman}\ and\ \citenamefont
  {Thampi}(2021)}]{purushothaman2021hydrodynamic}%
  \BibitemOpen
  \bibfield  {author} {\bibinfo {author} {\bibfnamefont {A.}~\bibnamefont
  {Purushothaman}}\ and\ \bibinfo {author} {\bibfnamefont {S.~P.}\ \bibnamefont
  {Thampi}},\ }\href {\doibase 10.1039/D0SM02140G} {\bibfield  {journal}
  {\bibinfo  {journal} {Soft Matter}\ } (\bibinfo {year} {2021}),\
  10.1039/D0SM02140G}\BibitemShut {NoStop}%
\bibitem [{\citenamefont {Ingham}\ \emph {et~al.}(2011)\citenamefont {Ingham},
  \citenamefont {Kalisman}, \citenamefont {Finkelshtein},\ and\ \citenamefont
  {Ben-Jacob}}]{ingham2011mutually}%
  \BibitemOpen
  \bibfield  {author} {\bibinfo {author} {\bibfnamefont {C.~J.}\ \bibnamefont
  {Ingham}}, \bibinfo {author} {\bibfnamefont {O.}~\bibnamefont {Kalisman}},
  \bibinfo {author} {\bibfnamefont {A.}~\bibnamefont {Finkelshtein}}, \ and\
  \bibinfo {author} {\bibfnamefont {E.}~\bibnamefont {Ben-Jacob}},\ }\href
  {\doibase 10.1073/pnas.1102097108} {\bibfield  {journal} {\bibinfo  {journal}
  {Proc. Nat. Acad. Sci.}\ }\textbf {\bibinfo {volume} {108}},\ \bibinfo
  {pages} {19731} (\bibinfo {year} {2011})}\BibitemShut {NoStop}%
\bibitem [{\citenamefont {Tuval}\ \emph {et~al.}(2005)\citenamefont {Tuval},
  \citenamefont {Cisneros}, \citenamefont {Dombrowski}, \citenamefont
  {Wolgemuth}, \citenamefont {Kessler},\ and\ \citenamefont
  {Goldstein}}]{tuval2005bacterial}%
  \BibitemOpen
  \bibfield  {author} {\bibinfo {author} {\bibfnamefont {I.}~\bibnamefont
  {Tuval}}, \bibinfo {author} {\bibfnamefont {L.}~\bibnamefont {Cisneros}},
  \bibinfo {author} {\bibfnamefont {C.}~\bibnamefont {Dombrowski}}, \bibinfo
  {author} {\bibfnamefont {C.~W.}\ \bibnamefont {Wolgemuth}}, \bibinfo {author}
  {\bibfnamefont {J.~O.}\ \bibnamefont {Kessler}}, \ and\ \bibinfo {author}
  {\bibfnamefont {R.~E.}\ \bibnamefont {Goldstein}},\ }\href {\doibase
  10.1073/pnas.0406724102} {\bibfield  {journal} {\bibinfo  {journal} {Proc.
  Nat. Acad. Sci.}\ }\textbf {\bibinfo {volume} {102}},\ \bibinfo {pages}
  {2277} (\bibinfo {year} {2005})}\BibitemShut {NoStop}%
\bibitem [{\citenamefont {Mathijssen}\ \emph
  {et~al.}(2018{\natexlab{b}})\citenamefont {Mathijssen}, \citenamefont
  {Jeanneret},\ and\ \citenamefont {Polin}}]{mathijssen2018_universal}%
  \BibitemOpen
  \bibfield  {author} {\bibinfo {author} {\bibfnamefont {A.~J. T.~M.}\
  \bibnamefont {Mathijssen}}, \bibinfo {author} {\bibfnamefont
  {R.}~\bibnamefont {Jeanneret}}, \ and\ \bibinfo {author} {\bibfnamefont
  {M.}~\bibnamefont {Polin}},\ }\href {\doibase 10.1103/PhysRevFluids.3.033103}
  {\bibfield  {journal} {\bibinfo  {journal} {Phys. Rev. Fluids}\ }\textbf
  {\bibinfo {volume} {3}},\ \bibinfo {pages} {033103} (\bibinfo {year}
  {2018}{\natexlab{b}})}\BibitemShut {NoStop}%
\bibitem [{\citenamefont {Dreyfus}\ \emph {et~al.}(2005)\citenamefont
  {Dreyfus}, \citenamefont {Baudry}, \citenamefont {Roper}, \citenamefont
  {Fermigier}, \citenamefont {Stone},\ and\ \citenamefont
  {Bibette}}]{dreyfus2005microscopic}%
  \BibitemOpen
  \bibfield  {author} {\bibinfo {author} {\bibfnamefont {R.}~\bibnamefont
  {Dreyfus}}, \bibinfo {author} {\bibfnamefont {J.}~\bibnamefont {Baudry}},
  \bibinfo {author} {\bibfnamefont {M.~L.}\ \bibnamefont {Roper}}, \bibinfo
  {author} {\bibfnamefont {M.}~\bibnamefont {Fermigier}}, \bibinfo {author}
  {\bibfnamefont {H.~A.}\ \bibnamefont {Stone}}, \ and\ \bibinfo {author}
  {\bibfnamefont {J.}~\bibnamefont {Bibette}},\ }\href {\doibase
  10.1038/nature04090} {\bibfield  {journal} {\bibinfo  {journal} {Nature}\
  }\textbf {\bibinfo {volume} {437}},\ \bibinfo {pages} {862} (\bibinfo {year}
  {2005})}\BibitemShut {NoStop}%
\bibitem [{\citenamefont {Howse}\ \emph {et~al.}(2007)\citenamefont {Howse},
  \citenamefont {Jones}, \citenamefont {Ryan}, \citenamefont {Gough},
  \citenamefont {Vafabakhsh},\ and\ \citenamefont
  {Golestanian}}]{howse2007selfmotile}%
  \BibitemOpen
  \bibfield  {author} {\bibinfo {author} {\bibfnamefont {J.~R.}\ \bibnamefont
  {Howse}}, \bibinfo {author} {\bibfnamefont {R.~A.~L.}\ \bibnamefont {Jones}},
  \bibinfo {author} {\bibfnamefont {A.~J.}\ \bibnamefont {Ryan}}, \bibinfo
  {author} {\bibfnamefont {T.}~\bibnamefont {Gough}}, \bibinfo {author}
  {\bibfnamefont {R.}~\bibnamefont {Vafabakhsh}}, \ and\ \bibinfo {author}
  {\bibfnamefont {R.}~\bibnamefont {Golestanian}},\ }\href {\doibase
  10.1103/PhysRevLett.99.048102} {\bibfield  {journal} {\bibinfo  {journal}
  {Phys. Rev. Lett.}\ }\textbf {\bibinfo {volume} {99}},\ \bibinfo {pages}
  {048102} (\bibinfo {year} {2007})}\BibitemShut {NoStop}%
\bibitem [{\citenamefont {Peng}\ \emph {et~al.}(2014)\citenamefont {Peng},
  \citenamefont {Lazo}, \citenamefont {Shiyanovskii},\ and\ \citenamefont
  {Lavrentovich}}]{peng2014induced}%
  \BibitemOpen
  \bibfield  {author} {\bibinfo {author} {\bibfnamefont {C.}~\bibnamefont
  {Peng}}, \bibinfo {author} {\bibfnamefont {I.}~\bibnamefont {Lazo}}, \bibinfo
  {author} {\bibfnamefont {S.~V.}\ \bibnamefont {Shiyanovskii}}, \ and\
  \bibinfo {author} {\bibfnamefont {O.~D.}\ \bibnamefont {Lavrentovich}},\
  }\href {\doibase 10.1103/PhysRevE.90.051002} {\bibfield  {journal} {\bibinfo
  {journal} {Phys. Rev. E}\ }\textbf {\bibinfo {volume} {90}},\ \bibinfo
  {pages} {051002(R)} (\bibinfo {year} {2014})}\BibitemShut {NoStop}%
\bibitem [{\citenamefont {Nishiguchi}\ and\ \citenamefont
  {Sano}(2015)}]{nishiguchi2015mesoscopic}%
  \BibitemOpen
  \bibfield  {author} {\bibinfo {author} {\bibfnamefont {D.}~\bibnamefont
  {Nishiguchi}}\ and\ \bibinfo {author} {\bibfnamefont {M.}~\bibnamefont
  {Sano}},\ }\href {\doibase 10.1103/PhysRevE.92.052309} {\bibfield  {journal}
  {\bibinfo  {journal} {Phys. Rev. E}\ }\textbf {\bibinfo {volume} {92}},\
  \bibinfo {pages} {052309} (\bibinfo {year} {2015})}\BibitemShut {NoStop}%
\bibitem [{\citenamefont {Bechinger}\ \emph {et~al.}(2016)\citenamefont
  {Bechinger}, \citenamefont {Di~Leonardo}, \citenamefont {L{\"o}wen},
  \citenamefont {Reichhardt}, \citenamefont {Volpe},\ and\ \citenamefont
  {Volpe}}]{bechinger2016active}%
  \BibitemOpen
  \bibfield  {author} {\bibinfo {author} {\bibfnamefont {C.}~\bibnamefont
  {Bechinger}}, \bibinfo {author} {\bibfnamefont {R.}~\bibnamefont
  {Di~Leonardo}}, \bibinfo {author} {\bibfnamefont {H.}~\bibnamefont
  {L{\"o}wen}}, \bibinfo {author} {\bibfnamefont {C.}~\bibnamefont
  {Reichhardt}}, \bibinfo {author} {\bibfnamefont {G.}~\bibnamefont {Volpe}}, \
  and\ \bibinfo {author} {\bibfnamefont {G.}~\bibnamefont {Volpe}},\ }\href
  {\doibase 10.1103/RevModPhys.88.045006} {\bibfield  {journal} {\bibinfo
  {journal} {Rev. Mod. Phys.}\ }\textbf {\bibinfo {volume} {88}},\ \bibinfo
  {pages} {045006} (\bibinfo {year} {2016})}\BibitemShut {NoStop}%
\bibitem [{\citenamefont {Elgeti}\ \emph {et~al.}(2015)\citenamefont {Elgeti},
  \citenamefont {Winkler},\ and\ \citenamefont {Gompper}}]{elgeti2015physics}%
  \BibitemOpen
  \bibfield  {author} {\bibinfo {author} {\bibfnamefont {J.}~\bibnamefont
  {Elgeti}}, \bibinfo {author} {\bibfnamefont {R.~G.}\ \bibnamefont {Winkler}},
  \ and\ \bibinfo {author} {\bibfnamefont {G.}~\bibnamefont {Gompper}},\ }\href
  {\doibase 10.1088/0034-4885/78/5/056601} {\bibfield  {journal} {\bibinfo
  {journal} {Rep. Progr. Phys.}\ }\textbf {\bibinfo {volume} {78}},\ \bibinfo
  {pages} {056601} (\bibinfo {year} {2015})}\BibitemShut {NoStop}%
\bibitem [{\citenamefont {Z{\"o}ttl}\ and\ \citenamefont
  {Stark}(2016)}]{zottl2016emergent}%
  \BibitemOpen
  \bibfield  {author} {\bibinfo {author} {\bibfnamefont {A.}~\bibnamefont
  {Z{\"o}ttl}}\ and\ \bibinfo {author} {\bibfnamefont {H.}~\bibnamefont
  {Stark}},\ }\href {\doibase 10.1088/0953-8984/28/25/253001} {\bibfield
  {journal} {\bibinfo  {journal} {J. Phys. Cond. Mat.}\ }\textbf {\bibinfo
  {volume} {28}},\ \bibinfo {pages} {253001} (\bibinfo {year}
  {2016})}\BibitemShut {NoStop}%
\bibitem [{\citenamefont {Campbell}\ \emph {et~al.}(2019)\citenamefont
  {Campbell}, \citenamefont {Ebbens}, \citenamefont {Illien},\ and\
  \citenamefont {Golestanian}}]{campbell2019_experimental}%
  \BibitemOpen
  \bibfield  {author} {\bibinfo {author} {\bibfnamefont {A.~I.}\ \bibnamefont
  {Campbell}}, \bibinfo {author} {\bibfnamefont {S.~J.}\ \bibnamefont
  {Ebbens}}, \bibinfo {author} {\bibfnamefont {P.}~\bibnamefont {Illien}}, \
  and\ \bibinfo {author} {\bibfnamefont {R.}~\bibnamefont {Golestanian}},\
  }\href {\doibase 10.1038/s41467-019-11842-1} {\bibfield  {journal} {\bibinfo
  {journal} {Nat. Commun.}\ }\textbf {\bibinfo {volume} {10}},\ \bibinfo
  {pages} {1} (\bibinfo {year} {2019})}\BibitemShut {NoStop}%
\bibitem [{\citenamefont {Nelson}\ \emph {et~al.}(2010)\citenamefont {Nelson},
  \citenamefont {Kaliakatsos},\ and\ \citenamefont
  {Abbott}}]{nelson2010microrobots}%
  \BibitemOpen
  \bibfield  {author} {\bibinfo {author} {\bibfnamefont {B.~J.}\ \bibnamefont
  {Nelson}}, \bibinfo {author} {\bibfnamefont {I.~K.}\ \bibnamefont
  {Kaliakatsos}}, \ and\ \bibinfo {author} {\bibfnamefont {J.~J.}\ \bibnamefont
  {Abbott}},\ }\href {\doibase 10.1146/annurev-bioeng-010510-103409} {\bibfield
   {journal} {\bibinfo  {journal} {Annu. Rev. Biomed. Eng.}\ }\textbf {\bibinfo
  {volume} {12}},\ \bibinfo {pages} {55} (\bibinfo {year} {2010})}\BibitemShut
  {NoStop}%
\bibitem [{\citenamefont {Tiwari}\ \emph {et~al.}(2012)\citenamefont {Tiwari},
  \citenamefont {Tiwari}, \citenamefont {Sriwastawa}, \citenamefont {Bhati},
  \citenamefont {Pandey}, \citenamefont {Pandey},\ and\ \citenamefont
  {Bannerjee}}]{tiwari2012drug}%
  \BibitemOpen
  \bibfield  {author} {\bibinfo {author} {\bibfnamefont {G.}~\bibnamefont
  {Tiwari}}, \bibinfo {author} {\bibfnamefont {R.}~\bibnamefont {Tiwari}},
  \bibinfo {author} {\bibfnamefont {B.}~\bibnamefont {Sriwastawa}}, \bibinfo
  {author} {\bibfnamefont {L.}~\bibnamefont {Bhati}}, \bibinfo {author}
  {\bibfnamefont {S.}~\bibnamefont {Pandey}}, \bibinfo {author} {\bibfnamefont
  {P.}~\bibnamefont {Pandey}}, \ and\ \bibinfo {author} {\bibfnamefont {S.~K.}\
  \bibnamefont {Bannerjee}},\ }\href {\doibase 10.4103/2230-973X.96920}
  {\bibfield  {journal} {\bibinfo  {journal} {Int. J. Pharmac. Inves.}\
  }\textbf {\bibinfo {volume} {2}},\ \bibinfo {pages} {2} (\bibinfo {year}
  {2012})}\BibitemShut {NoStop}%
\bibitem [{\citenamefont {Wilhelm}\ \emph {et~al.}(2016)\citenamefont
  {Wilhelm}, \citenamefont {Tavares}, \citenamefont {Dai}, \citenamefont
  {Ohta}, \citenamefont {Audet}, \citenamefont {Dvorak},\ and\ \citenamefont
  {Chan}}]{wilhelm2016analysis}%
  \BibitemOpen
  \bibfield  {author} {\bibinfo {author} {\bibfnamefont {S.}~\bibnamefont
  {Wilhelm}}, \bibinfo {author} {\bibfnamefont {A.~J.}\ \bibnamefont
  {Tavares}}, \bibinfo {author} {\bibfnamefont {Q.}~\bibnamefont {Dai}},
  \bibinfo {author} {\bibfnamefont {S.}~\bibnamefont {Ohta}}, \bibinfo {author}
  {\bibfnamefont {J.}~\bibnamefont {Audet}}, \bibinfo {author} {\bibfnamefont
  {H.~F.}\ \bibnamefont {Dvorak}}, \ and\ \bibinfo {author} {\bibfnamefont
  {W.~C.}\ \bibnamefont {Chan}},\ }\href {\doibase 10.1038/natrevmats.2016.14}
  {\bibfield  {journal} {\bibinfo  {journal} {Nat. Rev. Mater.}\ }\textbf
  {\bibinfo {volume} {1}},\ \bibinfo {pages} {1} (\bibinfo {year}
  {2016})}\BibitemShut {NoStop}%
\bibitem [{\citenamefont {Shen}\ \emph {et~al.}(2017)\citenamefont {Shen},
  \citenamefont {Wu}, \citenamefont {Liu},\ and\ \citenamefont
  {Wu}}]{shen2017high}%
  \BibitemOpen
  \bibfield  {author} {\bibinfo {author} {\bibfnamefont {S.}~\bibnamefont
  {Shen}}, \bibinfo {author} {\bibfnamefont {Y.}~\bibnamefont {Wu}}, \bibinfo
  {author} {\bibfnamefont {Y.}~\bibnamefont {Liu}}, \ and\ \bibinfo {author}
  {\bibfnamefont {D.}~\bibnamefont {Wu}},\ }\href {\doibase
  10.2147/IJN.S132780} {\bibfield  {journal} {\bibinfo  {journal} {Int. J.
  Nanomed.}\ }\textbf {\bibinfo {volume} {12}},\ \bibinfo {pages} {4085}
  (\bibinfo {year} {2017})}\BibitemShut {NoStop}%
\bibitem [{\citenamefont {Xu}\ \emph {et~al.}(2019)\citenamefont {Xu},
  \citenamefont {Wang}, \citenamefont {Liang}, \citenamefont {You},
  \citenamefont {Sanchez},\ and\ \citenamefont {Ma}}]{xu2019self}%
  \BibitemOpen
  \bibfield  {author} {\bibinfo {author} {\bibfnamefont {D.}~\bibnamefont
  {Xu}}, \bibinfo {author} {\bibfnamefont {Y.}~\bibnamefont {Wang}}, \bibinfo
  {author} {\bibfnamefont {C.}~\bibnamefont {Liang}}, \bibinfo {author}
  {\bibfnamefont {Y.}~\bibnamefont {You}}, \bibinfo {author} {\bibfnamefont
  {S.}~\bibnamefont {Sanchez}}, \ and\ \bibinfo {author} {\bibfnamefont
  {X.}~\bibnamefont {Ma}},\ }\href {\doibase 10.1002/smll.201902464} {\bibfield
   {journal} {\bibinfo  {journal} {Small}\ }\textbf {\bibinfo {volume} {16}},\
  \bibinfo {pages} {1902464} (\bibinfo {year} {2019})}\BibitemShut {NoStop}%
\bibitem [{\citenamefont {Alapan}\ \emph {et~al.}(2019)\citenamefont {Alapan},
  \citenamefont {Yasa}, \citenamefont {Yigit}, \citenamefont {Yasa},
  \citenamefont {Erkoc},\ and\ \citenamefont
  {Sitti}}]{alapan2019microrobotics}%
  \BibitemOpen
  \bibfield  {author} {\bibinfo {author} {\bibfnamefont {Y.}~\bibnamefont
  {Alapan}}, \bibinfo {author} {\bibfnamefont {O.}~\bibnamefont {Yasa}},
  \bibinfo {author} {\bibfnamefont {B.}~\bibnamefont {Yigit}}, \bibinfo
  {author} {\bibfnamefont {I.~C.}\ \bibnamefont {Yasa}}, \bibinfo {author}
  {\bibfnamefont {P.}~\bibnamefont {Erkoc}}, \ and\ \bibinfo {author}
  {\bibfnamefont {M.}~\bibnamefont {Sitti}},\ }\href {\doibase
  10.1146/annurev-control-053018-023803} {\bibfield  {journal} {\bibinfo
  {journal} {Annu. Rev. Control Robot. Autonom. Sys.}\ }\textbf {\bibinfo
  {volume} {2}},\ \bibinfo {pages} {205} (\bibinfo {year} {2019})}\BibitemShut
  {NoStop}%
\bibitem [{\citenamefont {Erkoc}\ \emph {et~al.}(2019)\citenamefont {Erkoc},
  \citenamefont {Yasa}, \citenamefont {Ceylan}, \citenamefont {Yasa},
  \citenamefont {Alapan},\ and\ \citenamefont {Sitti}}]{erkoc2019mobile}%
  \BibitemOpen
  \bibfield  {author} {\bibinfo {author} {\bibfnamefont {P.}~\bibnamefont
  {Erkoc}}, \bibinfo {author} {\bibfnamefont {I.~C.}\ \bibnamefont {Yasa}},
  \bibinfo {author} {\bibfnamefont {H.}~\bibnamefont {Ceylan}}, \bibinfo
  {author} {\bibfnamefont {O.}~\bibnamefont {Yasa}}, \bibinfo {author}
  {\bibfnamefont {Y.}~\bibnamefont {Alapan}}, \ and\ \bibinfo {author}
  {\bibfnamefont {M.}~\bibnamefont {Sitti}},\ }\href {\doibase
  10.1002/adtp.201800064} {\bibfield  {journal} {\bibinfo  {journal} {Adv.
  Therap.}\ }\textbf {\bibinfo {volume} {2}},\ \bibinfo {pages} {1800064}
  (\bibinfo {year} {2019})}\BibitemShut {NoStop}%
\bibitem [{\citenamefont {Daddi-Moussa-Ider}\ \emph {et~al.}(2020)\citenamefont
  {Daddi-Moussa-Ider}, \citenamefont {Lisicki},\ and\ \citenamefont
  {Mathijssen}}]{ider2020tuning}%
  \BibitemOpen
  \bibfield  {author} {\bibinfo {author} {\bibfnamefont {A.}~\bibnamefont
  {Daddi-Moussa-Ider}}, \bibinfo {author} {\bibfnamefont {M.}~\bibnamefont
  {Lisicki}}, \ and\ \bibinfo {author} {\bibfnamefont {A.~J. T.~M.}\
  \bibnamefont {Mathijssen}},\ }\href {\doibase
  10.1103/PhysRevApplied.14.024071} {\bibfield  {journal} {\bibinfo  {journal}
  {Phys. Rev. Applied}\ }\textbf {\bibinfo {volume} {14}},\ \bibinfo {pages}
  {024071} (\bibinfo {year} {2020})}\BibitemShut {NoStop}%
\bibitem [{\citenamefont {Yang}\ and\ \citenamefont
  {Zhang}(2020)}]{yang2020motion}%
  \BibitemOpen
  \bibfield  {author} {\bibinfo {author} {\bibfnamefont {L.}~\bibnamefont
  {Yang}}\ and\ \bibinfo {author} {\bibfnamefont {L.}~\bibnamefont {Zhang}},\
  }\href {\doibase 10.1146/annurev-control-032720-104318} {\bibfield  {journal}
  {\bibinfo  {journal} {Annu. Rev. Control Robot. Autonom. Sys.}\ }\textbf
  {\bibinfo {volume} {4}},\ \bibinfo {pages} {1.1–1.26} (\bibinfo {year}
  {2020})}\BibitemShut {NoStop}%
\bibitem [{\citenamefont {Thutupalli}\ \emph {et~al.}(2011)\citenamefont
  {Thutupalli}, \citenamefont {Seemann},\ and\ \citenamefont
  {Herminghaus}}]{thutupalli2011swarming}%
  \BibitemOpen
  \bibfield  {author} {\bibinfo {author} {\bibfnamefont {S.}~\bibnamefont
  {Thutupalli}}, \bibinfo {author} {\bibfnamefont {R.}~\bibnamefont {Seemann}},
  \ and\ \bibinfo {author} {\bibfnamefont {S.}~\bibnamefont {Herminghaus}},\
  }\href {\doibase 10.1088/1367-2630/13/7/073021} {\bibfield  {journal}
  {\bibinfo  {journal} {New J. Phys.}\ }\textbf {\bibinfo {volume} {13}},\
  \bibinfo {pages} {073021} (\bibinfo {year} {2011})}\BibitemShut {NoStop}%
\bibitem [{\citenamefont {Maass}\ \emph {et~al.}(2016)\citenamefont {Maass},
  \citenamefont {Kr{\"u}ger}, \citenamefont {Herminghaus},\ and\ \citenamefont
  {Bahr}}]{maass2016_swimming}%
  \BibitemOpen
  \bibfield  {author} {\bibinfo {author} {\bibfnamefont {C.~C.}\ \bibnamefont
  {Maass}}, \bibinfo {author} {\bibfnamefont {C.}~\bibnamefont {Kr{\"u}ger}},
  \bibinfo {author} {\bibfnamefont {S.}~\bibnamefont {Herminghaus}}, \ and\
  \bibinfo {author} {\bibfnamefont {C.}~\bibnamefont {Bahr}},\ }\href {\doibase
  10.1146/annurev-conmatphys-031115-011517} {\bibfield  {journal} {\bibinfo
  {journal} {Annu. Rev. Condens. Matter Phys.}\ }\textbf {\bibinfo {volume}
  {7}},\ \bibinfo {pages} {171} (\bibinfo {year} {2016})}\BibitemShut {NoStop}%
\bibitem [{\citenamefont {Thutupalli}\ and\ \citenamefont
  {Herminghaus}(2013)}]{thutupalli2013tuning}%
  \BibitemOpen
  \bibfield  {author} {\bibinfo {author} {\bibfnamefont {S.}~\bibnamefont
  {Thutupalli}}\ and\ \bibinfo {author} {\bibfnamefont {S.}~\bibnamefont
  {Herminghaus}},\ }\href {\doibase 10.1140/epje/i2013-13091-2} {\bibfield
  {journal} {\bibinfo  {journal} {Eur. Phys. J. E}\ }\textbf {\bibinfo {volume}
  {36}},\ \bibinfo {pages} {91} (\bibinfo {year} {2013})}\BibitemShut {NoStop}%
\bibitem [{\citenamefont {Izri}\ \emph {et~al.}(2014)\citenamefont {Izri},
  \citenamefont {van~der Linden}, \citenamefont {Michelin},\ and\ \citenamefont
  {Dauchot}}]{izri2014self}%
  \BibitemOpen
  \bibfield  {author} {\bibinfo {author} {\bibfnamefont {Z.}~\bibnamefont
  {Izri}}, \bibinfo {author} {\bibfnamefont {M.~N.}\ \bibnamefont {van~der
  Linden}}, \bibinfo {author} {\bibfnamefont {S.}~\bibnamefont {Michelin}}, \
  and\ \bibinfo {author} {\bibfnamefont {O.}~\bibnamefont {Dauchot}},\ }\href
  {\doibase 10.1103/PhysRevLett.113.248302} {\bibfield  {journal} {\bibinfo
  {journal} {Phys. Rev. Lett.}\ }\textbf {\bibinfo {volume} {113}},\ \bibinfo
  {pages} {248302} (\bibinfo {year} {2014})}\BibitemShut {NoStop}%
\bibitem [{\citenamefont {Krueger}\ \emph {et~al.}(2016)\citenamefont
  {Krueger}, \citenamefont {Bahr}, \citenamefont {Herminghaus},\ and\
  \citenamefont {Maass}}]{krueger2016dimensionality}%
  \BibitemOpen
  \bibfield  {author} {\bibinfo {author} {\bibfnamefont {C.}~\bibnamefont
  {Krueger}}, \bibinfo {author} {\bibfnamefont {C.}~\bibnamefont {Bahr}},
  \bibinfo {author} {\bibfnamefont {S.}~\bibnamefont {Herminghaus}}, \ and\
  \bibinfo {author} {\bibfnamefont {C.~C.}\ \bibnamefont {Maass}},\ }\href
  {\doibase 10.1140/epje/i2016-16064-y} {\bibfield  {journal} {\bibinfo
  {journal} {Eur. Phys. J. E}\ }\textbf {\bibinfo {volume} {39}},\ \bibinfo
  {pages} {64} (\bibinfo {year} {2016})}\BibitemShut {NoStop}%
\bibitem [{\citenamefont {Kr\"uger}\ \emph {et~al.}(2016)\citenamefont
  {Kr\"uger}, \citenamefont {Kl\"os}, \citenamefont {Bahr},\ and\ \citenamefont
  {Maass}}]{krueger2017curling}%
  \BibitemOpen
  \bibfield  {author} {\bibinfo {author} {\bibfnamefont {C.}~\bibnamefont
  {Kr\"uger}}, \bibinfo {author} {\bibfnamefont {G.}~\bibnamefont {Kl\"os}},
  \bibinfo {author} {\bibfnamefont {C.}~\bibnamefont {Bahr}}, \ and\ \bibinfo
  {author} {\bibfnamefont {C.~C.}\ \bibnamefont {Maass}},\ }\href {\doibase
  10.1103/PhysRevLett.117.048003} {\bibfield  {journal} {\bibinfo  {journal}
  {Phys. Rev. Lett.}\ }\textbf {\bibinfo {volume} {117}},\ \bibinfo {pages}
  {048003} (\bibinfo {year} {2016})}\BibitemShut {NoStop}%
\bibitem [{\citenamefont {Jin}\ \emph {et~al.}(2017)\citenamefont {Jin},
  \citenamefont {Kr{\"u}ger},\ and\ \citenamefont {Maass}}]{jin2017chemotaxis}%
  \BibitemOpen
  \bibfield  {author} {\bibinfo {author} {\bibfnamefont {C.}~\bibnamefont
  {Jin}}, \bibinfo {author} {\bibfnamefont {C.}~\bibnamefont {Kr{\"u}ger}}, \
  and\ \bibinfo {author} {\bibfnamefont {C.~C.}\ \bibnamefont {Maass}},\ }\href
  {\doibase 10.1073/pnas.1619783114} {\bibfield  {journal} {\bibinfo  {journal}
  {Proc. Nat. Acad. Sci.}\ }\textbf {\bibinfo {volume} {114}},\ \bibinfo
  {pages} {5089} (\bibinfo {year} {2017})}\BibitemShut {NoStop}%
\bibitem [{\citenamefont {Thutupalli}\ \emph {et~al.}(2018)\citenamefont
  {Thutupalli}, \citenamefont {Geyer}, \citenamefont {Singh}, \citenamefont
  {Adhikari},\ and\ \citenamefont {Stone}}]{thutupalli2018flow}%
  \BibitemOpen
  \bibfield  {author} {\bibinfo {author} {\bibfnamefont {S.}~\bibnamefont
  {Thutupalli}}, \bibinfo {author} {\bibfnamefont {D.}~\bibnamefont {Geyer}},
  \bibinfo {author} {\bibfnamefont {R.}~\bibnamefont {Singh}}, \bibinfo
  {author} {\bibfnamefont {R.}~\bibnamefont {Adhikari}}, \ and\ \bibinfo
  {author} {\bibfnamefont {H.~A.}\ \bibnamefont {Stone}},\ }\href {\doibase
  10.1073/pnas.1718807115} {\bibfield  {journal} {\bibinfo  {journal} {Proc.
  Nat. Acad. Sci.}\ }\textbf {\bibinfo {volume} {115}},\ \bibinfo {pages}
  {5403} (\bibinfo {year} {2018})}\BibitemShut {NoStop}%
\bibitem [{\citenamefont {Jin}\ \emph {et~al.}(2018)\citenamefont {Jin},
  \citenamefont {Hokmabad}, \citenamefont {Baldwin},\ and\ \citenamefont
  {Maass}}]{jin2018chemotactic}%
  \BibitemOpen
  \bibfield  {author} {\bibinfo {author} {\bibfnamefont {C.}~\bibnamefont
  {Jin}}, \bibinfo {author} {\bibfnamefont {B.~V.}\ \bibnamefont {Hokmabad}},
  \bibinfo {author} {\bibfnamefont {K.~A.}\ \bibnamefont {Baldwin}}, \ and\
  \bibinfo {author} {\bibfnamefont {C.~C.}\ \bibnamefont {Maass}},\ }\href
  {\doibase 10.1088/1361-648X/aaa208} {\bibfield  {journal} {\bibinfo
  {journal} {J. Phys. Cond. Matt.}\ }\textbf {\bibinfo {volume} {30}},\
  \bibinfo {pages} {054003} (\bibinfo {year} {2018})}\BibitemShut {NoStop}%
\bibitem [{\citenamefont {de~Blois}\ \emph {et~al.}(2019)\citenamefont
  {de~Blois}, \citenamefont {Reyssat}, \citenamefont {Michelin},\ and\
  \citenamefont {Dauchot}}]{blois2019flow}%
  \BibitemOpen
  \bibfield  {author} {\bibinfo {author} {\bibfnamefont {C.}~\bibnamefont
  {de~Blois}}, \bibinfo {author} {\bibfnamefont {M.}~\bibnamefont {Reyssat}},
  \bibinfo {author} {\bibfnamefont {S.}~\bibnamefont {Michelin}}, \ and\
  \bibinfo {author} {\bibfnamefont {O.}~\bibnamefont {Dauchot}},\ }\href
  {\doibase 10.1103/PhysRevFluids.4.054001} {\bibfield  {journal} {\bibinfo
  {journal} {Phys. Rev. Fluids}\ }\textbf {\bibinfo {volume} {4}},\ \bibinfo
  {pages} {054001} (\bibinfo {year} {2019})}\BibitemShut {NoStop}%
\bibitem [{\citenamefont {Stamatopoulos}\ \emph {et~al.}(2020)\citenamefont
  {Stamatopoulos}, \citenamefont {Milionis}, \citenamefont {Ackerl},
  \citenamefont {Donati}, \citenamefont {Leudet de~la Vall\'ee}, \citenamefont
  {Rudolf~von Rohr},\ and\ \citenamefont
  {Poulikakos}}]{stamatopoulos2020droplet}%
  \BibitemOpen
  \bibfield  {author} {\bibinfo {author} {\bibfnamefont {C.}~\bibnamefont
  {Stamatopoulos}}, \bibinfo {author} {\bibfnamefont {A.}~\bibnamefont
  {Milionis}}, \bibinfo {author} {\bibfnamefont {N.}~\bibnamefont {Ackerl}},
  \bibinfo {author} {\bibfnamefont {M.}~\bibnamefont {Donati}}, \bibinfo
  {author} {\bibfnamefont {P.}~\bibnamefont {Leudet de~la Vall\'ee}}, \bibinfo
  {author} {\bibfnamefont {P.}~\bibnamefont {Rudolf~von Rohr}}, \ and\ \bibinfo
  {author} {\bibfnamefont {D.}~\bibnamefont {Poulikakos}},\ }\href {\doibase
  10.1021/acsnano.0c03849} {\bibfield  {journal} {\bibinfo  {journal} {{ACS}
  Nano}\ }\textbf {\bibinfo {volume} {14}},\ \bibinfo {pages} {12895} (\bibinfo
  {year} {2020})}\BibitemShut {NoStop}%
\bibitem [{\citenamefont {Hokmabad}\ \emph {et~al.}(2021)\citenamefont
  {Hokmabad}, \citenamefont {Dey}, \citenamefont {Jalaal}, \citenamefont
  {Mohanty}, \citenamefont {Almukambetova}, \citenamefont {Baldwin},
  \citenamefont {Lohse},\ and\ \citenamefont {Maass}}]{hokmabad2021_emergence}%
  \BibitemOpen
  \bibfield  {author} {\bibinfo {author} {\bibfnamefont {B.~V.}\ \bibnamefont
  {Hokmabad}}, \bibinfo {author} {\bibfnamefont {R.}~\bibnamefont {Dey}},
  \bibinfo {author} {\bibfnamefont {M.}~\bibnamefont {Jalaal}}, \bibinfo
  {author} {\bibfnamefont {D.}~\bibnamefont {Mohanty}}, \bibinfo {author}
  {\bibfnamefont {M.}~\bibnamefont {Almukambetova}}, \bibinfo {author}
  {\bibfnamefont {K.~A.}\ \bibnamefont {Baldwin}}, \bibinfo {author}
  {\bibfnamefont {D.}~\bibnamefont {Lohse}}, \ and\ \bibinfo {author}
  {\bibfnamefont {C.~C.}\ \bibnamefont {Maass}},\ }\href {\doibase
  10.1103/PhysRevX.11.011043} {\bibfield  {journal} {\bibinfo  {journal} {Phys.
  Rev. X}\ }\textbf {\bibinfo {volume} {11}},\ \bibinfo {pages} {011043}
  (\bibinfo {year} {2021})}\BibitemShut {NoStop}%
\bibitem [{\citenamefont {Lippera}\ \emph {et~al.}(2020)\citenamefont
  {Lippera}, \citenamefont {Morozov}, \citenamefont {Benzaquen},\ and\
  \citenamefont {Michelin}}]{lippera2020collisions}%
  \BibitemOpen
  \bibfield  {author} {\bibinfo {author} {\bibfnamefont {K.}~\bibnamefont
  {Lippera}}, \bibinfo {author} {\bibfnamefont {M.}~\bibnamefont {Morozov}},
  \bibinfo {author} {\bibfnamefont {M.}~\bibnamefont {Benzaquen}}, \ and\
  \bibinfo {author} {\bibfnamefont {S.}~\bibnamefont {Michelin}},\ }\href
  {\doibase 10.1017/jfm.2019.1055} {\bibfield  {journal} {\bibinfo  {journal}
  {J. Fluid Mech.}\ }\textbf {\bibinfo {volume} {886}},\ \bibinfo {pages} {A17}
  (\bibinfo {year} {2020})}\BibitemShut {NoStop}%
\bibitem [{\citenamefont {Hokmabad}\ \emph {et~al.}(2020)\citenamefont
  {Hokmabad}, \citenamefont {Saha}, \citenamefont {Canalejo}, \citenamefont
  {Golestanian},\ and\ \citenamefont {Maass}}]{hokmabad2020quantitative}%
  \BibitemOpen
  \bibfield  {author} {\bibinfo {author} {\bibfnamefont {B.~V.}\ \bibnamefont
  {Hokmabad}}, \bibinfo {author} {\bibfnamefont {S.}~\bibnamefont {Saha}},
  \bibinfo {author} {\bibfnamefont {J.~A.}\ \bibnamefont {Canalejo}}, \bibinfo
  {author} {\bibfnamefont {R.}~\bibnamefont {Golestanian}}, \ and\ \bibinfo
  {author} {\bibfnamefont {C.~C.}\ \bibnamefont {Maass}},\ }\href
  {https://arxiv.org/abs/2012.05170} {\  (\bibinfo {year} {2020})},\ \Eprint
  {http://arxiv.org/abs/2012.05170} {arXiv:2012.05170} \BibitemShut {NoStop}%
\bibitem [{Note1()}]{Note1}%
  \BibitemOpen
  \bibinfo {note} {See Supplemental Material containing experimental methods,
  theoretical derivations, a supplementary figure, 3 videos, and additional
  references \cite {qin2010_soft, thorsen2001_dynamic, crocker1996_methods,
  Jeanneret2019, tsay1991viscous, pepper2010_nearby, Pushkin2016,
  nganguia2018squirming, nganguia2020squirming, whitaker1986flow,
  gilpin2017vortex, liu2016bimetallic, weeks1971role, eames1999connection,
  dabiri2005estimation}.}\BibitemShut {Stop}%
\bibitem [{\citenamefont {Herminghaus}\ \emph {et~al.}(2014)\citenamefont
  {Herminghaus}, \citenamefont {Maass}, \citenamefont {Kr{\"u}ger},
  \citenamefont {Thutupalli}, \citenamefont {Goehring},\ and\ \citenamefont
  {Bahr}}]{herminghaus2014interfacial}%
  \BibitemOpen
  \bibfield  {author} {\bibinfo {author} {\bibfnamefont {S.}~\bibnamefont
  {Herminghaus}}, \bibinfo {author} {\bibfnamefont {C.~C.}\ \bibnamefont
  {Maass}}, \bibinfo {author} {\bibfnamefont {C.}~\bibnamefont {Kr{\"u}ger}},
  \bibinfo {author} {\bibfnamefont {S.}~\bibnamefont {Thutupalli}}, \bibinfo
  {author} {\bibfnamefont {L.}~\bibnamefont {Goehring}}, \ and\ \bibinfo
  {author} {\bibfnamefont {C.}~\bibnamefont {Bahr}},\ }\href {\doibase
  10.1039/C4SM00550C} {\bibfield  {journal} {\bibinfo  {journal} {Soft Matter}\
  }\textbf {\bibinfo {volume} {10}},\ \bibinfo {pages} {7008} (\bibinfo {year}
  {2014})}\BibitemShut {NoStop}%
\bibitem [{\citenamefont {Lighthill}(1952)}]{lighthill1952squirming}%
  \BibitemOpen
  \bibfield  {author} {\bibinfo {author} {\bibfnamefont {M.~J.}\ \bibnamefont
  {Lighthill}},\ }\href {\doibase 10.1002/cpa.3160050201} {\bibfield  {journal}
  {\bibinfo  {journal} {Comm. Pure Appl. Math.}\ }\textbf {\bibinfo {volume}
  {5}},\ \bibinfo {pages} {109} (\bibinfo {year} {1952})}\BibitemShut {NoStop}%
\bibitem [{\citenamefont {Blake}(1971)}]{blake1971spherical}%
  \BibitemOpen
  \bibfield  {author} {\bibinfo {author} {\bibfnamefont {J.~R.}\ \bibnamefont
  {Blake}},\ }\href {\doibase 10.1017/S002211207100048X} {\bibfield  {journal}
  {\bibinfo  {journal} {J. Fluid Mech.}\ }\textbf {\bibinfo {volume} {46}},\
  \bibinfo {pages} {199} (\bibinfo {year} {1971})}\BibitemShut {NoStop}%
\bibitem [{\citenamefont {Brinkman}(1947)}]{brinkman1947_calculation}%
  \BibitemOpen
  \bibfield  {author} {\bibinfo {author} {\bibfnamefont {H.~C.}\ \bibnamefont
  {Brinkman}},\ }\href {\doibase 10.1016/0031-8914(47)90030-X} {\bibfield
  {journal} {\bibinfo  {journal} {Physica}\ }\textbf {\bibinfo {volume} {13}},\
  \bibinfo {pages} {447} (\bibinfo {year} {1947})}\BibitemShut {NoStop}%
\bibitem [{\citenamefont {Mathijssen}\ \emph {et~al.}(2015)\citenamefont
  {Mathijssen}, \citenamefont {Pushkin},\ and\ \citenamefont
  {Yeomans}}]{mathijssen2015_tracer}%
  \BibitemOpen
  \bibfield  {author} {\bibinfo {author} {\bibfnamefont {A.~J. T.~M.}\
  \bibnamefont {Mathijssen}}, \bibinfo {author} {\bibfnamefont {D.~O.}\
  \bibnamefont {Pushkin}}, \ and\ \bibinfo {author} {\bibfnamefont {J.~M.}\
  \bibnamefont {Yeomans}},\ }\href {\doibase 10.1017/jfm.2015.269} {\bibfield
  {journal} {\bibinfo  {journal} {J. Fluid Mech.}\ }\textbf {\bibinfo {volume}
  {773}},\ \bibinfo {pages} {498} (\bibinfo {year} {2015})}\BibitemShut
  {NoStop}%
\bibitem [{\citenamefont {Simmchen}\ \emph {et~al.}(2016)\citenamefont
  {Simmchen}, \citenamefont {Katuri}, \citenamefont {Uspal}, \citenamefont
  {Popescu}, \citenamefont {Tasinkevych},\ and\ \citenamefont
  {S{\'a}nchez}}]{simmchen2016topographical}%
  \BibitemOpen
  \bibfield  {author} {\bibinfo {author} {\bibfnamefont {J.}~\bibnamefont
  {Simmchen}}, \bibinfo {author} {\bibfnamefont {J.}~\bibnamefont {Katuri}},
  \bibinfo {author} {\bibfnamefont {W.~E.}\ \bibnamefont {Uspal}}, \bibinfo
  {author} {\bibfnamefont {M.~N.}\ \bibnamefont {Popescu}}, \bibinfo {author}
  {\bibfnamefont {M.}~\bibnamefont {Tasinkevych}}, \ and\ \bibinfo {author}
  {\bibfnamefont {S.}~\bibnamefont {S{\'a}nchez}},\ }\href {\doibase
  10.1038/ncomms10598} {\bibfield  {journal} {\bibinfo  {journal} {Nat.
  Commun.}\ }\textbf {\bibinfo {volume} {7}},\ \bibinfo {pages} {10598}
  (\bibinfo {year} {2016})}\BibitemShut {NoStop}%
\bibitem [{\citenamefont {Z{\"o}ttl}\ and\ \citenamefont
  {Stark}(2014)}]{zottl2014hydrodynamics}%
  \BibitemOpen
  \bibfield  {author} {\bibinfo {author} {\bibfnamefont {A.}~\bibnamefont
  {Z{\"o}ttl}}\ and\ \bibinfo {author} {\bibfnamefont {H.}~\bibnamefont
  {Stark}},\ }\href {\doibase 10.1103/PhysRevLett.112.118101} {\bibfield
  {journal} {\bibinfo  {journal} {Phys. Rev. Lett.}\ }\textbf {\bibinfo
  {volume} {112}},\ \bibinfo {pages} {118101} (\bibinfo {year}
  {2014})}\BibitemShut {NoStop}%
\bibitem [{\citenamefont {Wioland}\ \emph {et~al.}(2016)\citenamefont
  {Wioland}, \citenamefont {Lushi},\ and\ \citenamefont
  {Goldstein}}]{wioland2016directed}%
  \BibitemOpen
  \bibfield  {author} {\bibinfo {author} {\bibfnamefont {H.}~\bibnamefont
  {Wioland}}, \bibinfo {author} {\bibfnamefont {E.}~\bibnamefont {Lushi}}, \
  and\ \bibinfo {author} {\bibfnamefont {R.~E.}\ \bibnamefont {Goldstein}},\
  }\href {\doibase 10.1088/1367-2630/18/7/075002} {\bibfield  {journal}
  {\bibinfo  {journal} {New J. Phys.}\ }\textbf {\bibinfo {volume} {18}},\
  \bibinfo {pages} {075002} (\bibinfo {year} {2016})}\BibitemShut {NoStop}%
\bibitem [{\citenamefont {Vidakovic}\ \emph {et~al.}(2018)\citenamefont
  {Vidakovic}, \citenamefont {Singh}, \citenamefont {Hartmann}, \citenamefont
  {Nadell},\ and\ \citenamefont {Drescher}}]{vidakovic2018dynamic}%
  \BibitemOpen
  \bibfield  {author} {\bibinfo {author} {\bibfnamefont {L.}~\bibnamefont
  {Vidakovic}}, \bibinfo {author} {\bibfnamefont {P.~K.}\ \bibnamefont
  {Singh}}, \bibinfo {author} {\bibfnamefont {R.}~\bibnamefont {Hartmann}},
  \bibinfo {author} {\bibfnamefont {C.~D.}\ \bibnamefont {Nadell}}, \ and\
  \bibinfo {author} {\bibfnamefont {K.}~\bibnamefont {Drescher}},\ }\href
  {\doibase 10.1038/s41564-017-0050-1} {\bibfield  {journal} {\bibinfo
  {journal} {Nat. Microbiol.}\ }\textbf {\bibinfo {volume} {3}},\ \bibinfo
  {pages} {26} (\bibinfo {year} {2018})}\BibitemShut {NoStop}%
\bibitem [{\citenamefont {Bhattacharjee}\ and\ \citenamefont
  {Datta}(2019)}]{bhattacharjee2019bacterial}%
  \BibitemOpen
  \bibfield  {author} {\bibinfo {author} {\bibfnamefont {T.}~\bibnamefont
  {Bhattacharjee}}\ and\ \bibinfo {author} {\bibfnamefont {S.~S.}\ \bibnamefont
  {Datta}},\ }\href {\doibase 10.1038/s41467-019-10115-1} {\bibfield  {journal}
  {\bibinfo  {journal} {Nat. Commun.}\ }\textbf {\bibinfo {volume} {10}},\
  \bibinfo {pages} {1} (\bibinfo {year} {2019})}\BibitemShut {NoStop}%
\bibitem [{\citenamefont {Roveillo}\ \emph {et~al.}(2020)\citenamefont
  {Roveillo}, \citenamefont {Dervaux}, \citenamefont {Wang}, \citenamefont
  {Rouyer}, \citenamefont {Zanchi}, \citenamefont {Seuront},\ and\
  \citenamefont {Elias}}]{roveillo2020trapping}%
  \BibitemOpen
  \bibfield  {author} {\bibinfo {author} {\bibfnamefont {Q.}~\bibnamefont
  {Roveillo}}, \bibinfo {author} {\bibfnamefont {J.}~\bibnamefont {Dervaux}},
  \bibinfo {author} {\bibfnamefont {Y.}~\bibnamefont {Wang}}, \bibinfo {author}
  {\bibfnamefont {F.}~\bibnamefont {Rouyer}}, \bibinfo {author} {\bibfnamefont
  {D.}~\bibnamefont {Zanchi}}, \bibinfo {author} {\bibfnamefont
  {L.}~\bibnamefont {Seuront}}, \ and\ \bibinfo {author} {\bibfnamefont
  {F.}~\bibnamefont {Elias}},\ }\href {\doibase 10.1098/rsif.2020.0077}
  {\bibfield  {journal} {\bibinfo  {journal} {J. R. Soc. Interface}\ }\textbf
  {\bibinfo {volume} {17}},\ \bibinfo {pages} {20200077} (\bibinfo {year}
  {2020})}\BibitemShut {NoStop}%
\bibitem [{\citenamefont {Qin}\ \emph {et~al.}(2010)\citenamefont {Qin},
  \citenamefont {Xia},\ and\ \citenamefont {Whitesides}}]{qin2010_soft}%
  \BibitemOpen
  \bibfield  {author} {\bibinfo {author} {\bibfnamefont {D.}~\bibnamefont
  {Qin}}, \bibinfo {author} {\bibfnamefont {Y.}~\bibnamefont {Xia}}, \ and\
  \bibinfo {author} {\bibfnamefont {G.~M.}\ \bibnamefont {Whitesides}},\ }\href
  {\doibase 10.1038/nprot.2009.234} {\bibfield  {journal} {\bibinfo  {journal}
  {Nat. Protoc.}\ }\textbf {\bibinfo {volume} {5}},\ \bibinfo {pages} {491}
  (\bibinfo {year} {2010})}\BibitemShut {NoStop}%
\bibitem [{\citenamefont {Thorsen}\ \emph {et~al.}(2001)\citenamefont
  {Thorsen}, \citenamefont {Roberts}, \citenamefont {Arnold},\ and\
  \citenamefont {Quake}}]{thorsen2001_dynamic}%
  \BibitemOpen
  \bibfield  {author} {\bibinfo {author} {\bibfnamefont {T.}~\bibnamefont
  {Thorsen}}, \bibinfo {author} {\bibfnamefont {R.~W.}\ \bibnamefont
  {Roberts}}, \bibinfo {author} {\bibfnamefont {F.~H.}\ \bibnamefont {Arnold}},
  \ and\ \bibinfo {author} {\bibfnamefont {S.~R.}\ \bibnamefont {Quake}},\
  }\href {\doibase 10.1103/PhysRevLett.86.4163} {\bibfield  {journal} {\bibinfo
   {journal} {Phys. Rev. Lett.}\ }\textbf {\bibinfo {volume} {86}},\ \bibinfo
  {pages} {4163} (\bibinfo {year} {2001})}\BibitemShut {NoStop}%
\bibitem [{\citenamefont {Crocker}\ and\ \citenamefont
  {Grier}(1996)}]{crocker1996_methods}%
  \BibitemOpen
  \bibfield  {author} {\bibinfo {author} {\bibfnamefont {J.~C.}\ \bibnamefont
  {Crocker}}\ and\ \bibinfo {author} {\bibfnamefont {D.~G.}\ \bibnamefont
  {Grier}},\ }\href {\doibase 10.1006/jcis.1996.0217} {\bibfield  {journal}
  {\bibinfo  {journal} {J. Colloid Interface Sci.}\ }\textbf {\bibinfo {volume}
  {179}},\ \bibinfo {pages} {298} (\bibinfo {year} {1996})}\BibitemShut
  {NoStop}%
\bibitem [{\citenamefont {Jeanneret}\ \emph {et~al.}(2019)\citenamefont
  {Jeanneret}, \citenamefont {Pushkin},\ and\ \citenamefont
  {Polin}}]{Jeanneret2019}%
  \BibitemOpen
  \bibfield  {author} {\bibinfo {author} {\bibfnamefont {R.}~\bibnamefont
  {Jeanneret}}, \bibinfo {author} {\bibfnamefont {D.~O.}\ \bibnamefont
  {Pushkin}}, \ and\ \bibinfo {author} {\bibfnamefont {M.}~\bibnamefont
  {Polin}},\ }\href {\doibase 10.1103/PhysRevLett.123.248102} {\bibfield
  {journal} {\bibinfo  {journal} {Phys. Rev. Lett.}\ }\textbf {\bibinfo
  {volume} {123}},\ \bibinfo {pages} {248102} (\bibinfo {year}
  {2019})}\BibitemShut {NoStop}%
\bibitem [{\citenamefont {Tsay}\ and\ \citenamefont
  {Weinbaum}(1991)}]{tsay1991viscous}%
  \BibitemOpen
  \bibfield  {author} {\bibinfo {author} {\bibfnamefont {R.-Y.}\ \bibnamefont
  {Tsay}}\ and\ \bibinfo {author} {\bibfnamefont {S.}~\bibnamefont
  {Weinbaum}},\ }\href {\doibase 10.1017/S0022112091002318} {\bibfield
  {journal} {\bibinfo  {journal} {J. Fluid Mech.}\ }\textbf {\bibinfo {volume}
  {226}},\ \bibinfo {pages} {125} (\bibinfo {year} {1991})}\BibitemShut
  {NoStop}%
\bibitem [{\citenamefont {Pepper}\ \emph {et~al.}(2010)\citenamefont {Pepper},
  \citenamefont {Roper}, \citenamefont {Ryu}, \citenamefont {Matsudaira},\ and\
  \citenamefont {Stone}}]{pepper2010_nearby}%
  \BibitemOpen
  \bibfield  {author} {\bibinfo {author} {\bibfnamefont {R.~E.}\ \bibnamefont
  {Pepper}}, \bibinfo {author} {\bibfnamefont {M.}~\bibnamefont {Roper}},
  \bibinfo {author} {\bibfnamefont {S.}~\bibnamefont {Ryu}}, \bibinfo {author}
  {\bibfnamefont {P.}~\bibnamefont {Matsudaira}}, \ and\ \bibinfo {author}
  {\bibfnamefont {H.~A.}\ \bibnamefont {Stone}},\ }\href {\doibase
  10.1098/rsif.2009.0419} {\bibfield  {journal} {\bibinfo  {journal} {J. R.
  Soc. Interface}\ }\textbf {\bibinfo {volume} {7}},\ \bibinfo {pages} {851}
  (\bibinfo {year} {2010})}\BibitemShut {NoStop}%
\bibitem [{\citenamefont {Pushkin}\ and\ \citenamefont
  {Bees}(2016)}]{Pushkin2016}%
  \BibitemOpen
  \bibfield  {author} {\bibinfo {author} {\bibfnamefont {D.~O.}\ \bibnamefont
  {Pushkin}}\ and\ \bibinfo {author} {\bibfnamefont {M.~A.}\ \bibnamefont
  {Bees}},\ }in\ \href {\doibase 10.1007/978-3-319-32189-9_12} {\emph {\bibinfo
  {booktitle} {Biophysics of Infection. Advances in Experimental Medicine and
  Biology}}},\ Vol.\ \bibinfo {volume} {915},\ \bibinfo {editor} {edited by\
  \bibinfo {editor} {\bibfnamefont {M.}~\bibnamefont {Leake}}}\ (\bibinfo
  {publisher} {Springer},\ \bibinfo {year} {2016})\ pp.\ \bibinfo {pages}
  {193--205}\BibitemShut {NoStop}%
\bibitem [{\citenamefont {Nganguia}\ and\ \citenamefont
  {Pak}(2018)}]{nganguia2018squirming}%
  \BibitemOpen
  \bibfield  {author} {\bibinfo {author} {\bibfnamefont {H.}~\bibnamefont
  {Nganguia}}\ and\ \bibinfo {author} {\bibfnamefont {O.~S.}\ \bibnamefont
  {Pak}},\ }\href {\doibase 10.1017/jfm.2018.685} {\bibfield  {journal}
  {\bibinfo  {journal} {J. Fluid Mech.}\ }\textbf {\bibinfo {volume} {855}},\
  \bibinfo {pages} {554} (\bibinfo {year} {2018})}\BibitemShut {NoStop}%
\bibitem [{\citenamefont {Nganguia}\ \emph {et~al.}(2020)\citenamefont
  {Nganguia}, \citenamefont {Zhu}, \citenamefont {Palaniappan},\ and\
  \citenamefont {Pak}}]{nganguia2020squirming}%
  \BibitemOpen
  \bibfield  {author} {\bibinfo {author} {\bibfnamefont {H.}~\bibnamefont
  {Nganguia}}, \bibinfo {author} {\bibfnamefont {L.}~\bibnamefont {Zhu}},
  \bibinfo {author} {\bibfnamefont {D.}~\bibnamefont {Palaniappan}}, \ and\
  \bibinfo {author} {\bibfnamefont {O.~S.}\ \bibnamefont {Pak}},\ }\href
  {\doibase 10.1103/PhysRevE.101.063105} {\bibfield  {journal} {\bibinfo
  {journal} {Phys. Rev. E}\ }\textbf {\bibinfo {volume} {101}},\ \bibinfo
  {pages} {063105} (\bibinfo {year} {2020})}\BibitemShut {NoStop}%
\bibitem [{\citenamefont {Whitaker}(1986)}]{whitaker1986flow}%
  \BibitemOpen
  \bibfield  {author} {\bibinfo {author} {\bibfnamefont {S.}~\bibnamefont
  {Whitaker}},\ }\href {\doibase 10.1007/BF01036523} {\bibfield  {journal}
  {\bibinfo  {journal} {Transport in porous media}\ }\textbf {\bibinfo {volume}
  {1}},\ \bibinfo {pages} {3} (\bibinfo {year} {1986})}\BibitemShut {NoStop}%
\bibitem [{\citenamefont {Gilpin}\ \emph {et~al.}(2017)\citenamefont {Gilpin},
  \citenamefont {Prakash},\ and\ \citenamefont {Prakash}}]{gilpin2017vortex}%
  \BibitemOpen
  \bibfield  {author} {\bibinfo {author} {\bibfnamefont {W.}~\bibnamefont
  {Gilpin}}, \bibinfo {author} {\bibfnamefont {V.~N.}\ \bibnamefont {Prakash}},
  \ and\ \bibinfo {author} {\bibfnamefont {M.}~\bibnamefont {Prakash}},\ }\href
  {\doibase 10.1038/nphys3981} {\bibfield  {journal} {\bibinfo  {journal} {Nat.
  Phys.}\ }\textbf {\bibinfo {volume} {13}},\ \bibinfo {pages} {380} (\bibinfo
  {year} {2017})}\BibitemShut {NoStop}%
\bibitem [{\citenamefont {Liu}\ \emph {et~al.}(2016)\citenamefont {Liu},
  \citenamefont {Zhou}, \citenamefont {Wang},\ and\ \citenamefont
  {Zhang}}]{liu2016bimetallic}%
  \BibitemOpen
  \bibfield  {author} {\bibinfo {author} {\bibfnamefont {C.}~\bibnamefont
  {Liu}}, \bibinfo {author} {\bibfnamefont {C.}~\bibnamefont {Zhou}}, \bibinfo
  {author} {\bibfnamefont {W.}~\bibnamefont {Wang}}, \ and\ \bibinfo {author}
  {\bibfnamefont {H.~P.}\ \bibnamefont {Zhang}},\ }\href {\doibase
  10.1103/PhysRevLett.117.198001} {\bibfield  {journal} {\bibinfo  {journal}
  {Phys. Rev. Lett.}\ }\textbf {\bibinfo {volume} {117}},\ \bibinfo {pages}
  {198001} (\bibinfo {year} {2016})}\BibitemShut {NoStop}%
\bibitem [{\citenamefont {Weeks}\ \emph {et~al.}(1971)\citenamefont {Weeks},
  \citenamefont {Chandler},\ and\ \citenamefont {Andersen}}]{weeks1971role}%
  \BibitemOpen
  \bibfield  {author} {\bibinfo {author} {\bibfnamefont {J.~D.}\ \bibnamefont
  {Weeks}}, \bibinfo {author} {\bibfnamefont {D.}~\bibnamefont {Chandler}}, \
  and\ \bibinfo {author} {\bibfnamefont {H.~C.}\ \bibnamefont {Andersen}},\
  }\href {\doibase 10.1063/1.1674820} {\bibfield  {journal} {\bibinfo
  {journal} {J. Chem. Phys.}\ }\textbf {\bibinfo {volume} {54}},\ \bibinfo
  {pages} {5237} (\bibinfo {year} {1971})}\BibitemShut {NoStop}%
\bibitem [{\citenamefont {Eames}\ and\ \citenamefont
  {McIntyre}(1999)}]{eames1999connection}%
  \BibitemOpen
  \bibfield  {author} {\bibinfo {author} {\bibfnamefont {I.}~\bibnamefont
  {Eames}}\ and\ \bibinfo {author} {\bibfnamefont {M.~E.}\ \bibnamefont
  {McIntyre}},\ }\href {\doibase 10.1017/S0305004198003223} {\bibfield
  {journal} {\bibinfo  {journal} {Math. Proc. Cambridge Phil. Soc.}\ }\textbf
  {\bibinfo {volume} {126}},\ \bibinfo {pages} {171} (\bibinfo {year}
  {1999})}\BibitemShut {NoStop}%
\bibitem [{\citenamefont {Dabiri}(2005)}]{dabiri2005estimation}%
  \BibitemOpen
  \bibfield  {author} {\bibinfo {author} {\bibfnamefont {J.~O.}\ \bibnamefont
  {Dabiri}},\ }\href {\doibase 10.1242/jeb.01813} {\bibfield  {journal}
  {\bibinfo  {journal} {J. Exp. Biol.}\ }\textbf {\bibinfo {volume} {208}},\
  \bibinfo {pages} {3519} (\bibinfo {year} {2005})}\BibitemShut {NoStop}%
\end{thebibliography}%
